\documentclass[a4paper,aps,onecolumn,nofootinbib]{revtex4}
\RequirePackage[colorlinks,hyperindex]{hyperref}
\RequirePackage[english]{babel}
\RequirePackage[latin1]{inputenc}
\RequirePackage[T1]{fontenc}
\RequirePackage{mathrsfs}
\RequirePackage{amsmath}
\RequirePackage{amssymb}
\RequirePackage{amsbsy}
\RequirePackage{color}
\RequirePackage{bm}
\hypersetup{colorlinks=true,breaklinks=true,urlcolor=blue,linkcolor=red}
\begin{document}
%%%%%%%%%%%%%%%%%%%%%%%%%%%%%%%%%%%%%%%%%%%%%%%%%%%%%%%%%%%%%%%%%%%%%%%%%%%%%%%%%%%%%%%%%%%%%%%%%%%
\title{\bf{Dirac Fields in Hydrodynamic Form and their Thermodynamic Formulation}}
\author{Luca Fabbri,$^{c}$\!\! $^{\hbar}$\!\! $^{G}$\footnote{luca.fabbri@unige.it}
Stefano Vignolo,$^{c}$\!\! $^{\hbar}$\!\! $^{G}$\footnote{stefano.vignolo@unige.it}
Giuseppe De Maria,$^{c}$\!\! $^{\hbar}$\footnote{giuseppe.demaria@edu.unige.it}
Sante Carloni,$^{c}$\!\! $^{\hbar}$\!\! $^{G}$\!\! $^{\nabla}$\footnote{sante.carloni@unige.it}}
\affiliation{$^{c}$DIME, Universit\`{a} di Genova, Via all'Opera Pia 15, 16145 Genova, ITALY\\
$^{\hbar}$INFN, Sezione di Genova, Via Dodecaneso 33, 16146 Genova, ITALY\\
$^{G}$GNFM, Istituto Nazionale di Alta Matematica, P.le Aldo Moro 5, 00185 Roma, ITALY\\
$^{\nabla}$Institute of Theoretical Physics, Faculty of Mathematics and Physics,\\
Charles University, Prague, V Hole\v{s}ovi\v{c}k\'{a}ch 2, 180 00 Prague 8, CZECH REPUBLIC}
\date{\today}
%%%%%%%%%%%%%%%%%%%%%%%%%%%%%%%%%%%%%%%%%%%%%%%%%%%%%%%%%%%%%%%%%%%%%%%%%%%%%%%%%%%%%%%%%%%%%%%%%%%
\begin{abstract}
We consider the theory of spinor fields written in polar form and we re-express it in terms of the so-called $1\!+\!1\!+\!2$ covariant splitting: after this is done for the basic kinematic variables, we proceed to decompose the dynamical equations, both for the case of the Dirac differential field equations and for the case of the energy density tensor. As an explicit example of a real physical application we deal with the hydrogen atom, superconductivity and an analogy with the van der Waals gas.
\end{abstract}
%%%%%%%%%%%%%%%%%%%%%%%%%%%%%%%%%%%%%%%%%%%%%%%%%%%%%%%%%%%%%%%%%%%%%%%%%%%%%%%%%%%%%%%%%%%%%%%%%%%
\maketitle
%%%%%%%%%%%%%%%%%%%%%%%%%%%%%%%%%%%%%%%%%%%%%%%%%%%%%%%%%%%%%%%%%%%%%%%%%%%%%%%%%%%%%%%%%%%%%%%%%%%
\section{Introduction}
In mathematics, a complex function is said to have \emph{polar form}\footnote{It is important to specify here that the adjective 'polar' will always be used in concomitance to the decomposition of complex objects into real objects, and never in concomitance with cylindrical coordinates (which we will never employ).} when it is expressed as the product of a module and a unitary phase. In physics, this can be done whenever a physical object is described in terms of complex functions, and in particular, for spinor fields in relativistic quantum mechanics \cite{jl1}. When the relativistic spinor field is re-configured in polar form, the corresponding Dirac theory is re-arranged as a type of \emph{hydrodynamics}\footnote{It is also important to clarify that the term 'hydrodynamics' was chosen a century ago as a metaphor to describe something that behaves like some sort of incompressible fluid, not because it was believed that there could be any resemblance to water.} \cite{t2}.

Writing spinors in polar form has several advantages: 1. the theory is re-formulated only with real variables; 2. the specific representation of the gamma matrices becomes irrelevant, because no gamma matrices appear; 3. tetrads and co-tetrads are no longer needed to fix the soldering between the tensor algebra and the geometry of the spacetime; 4. it is manifestly covariant and so applicable to any system of coordinates, in particular in curved spacetimes; 5. other interactions like electrodynamics are also automatically incorporated; 6. the hydrodynamic equations are classical in form, and therefore they can be studied by means of the methods of differential geometry and fluid dynamics \cite{Fabbri:2023onb, Fabbri:2023yhl, Fabbri:2025ftm, Fabbri:2024avj}.

With this as basis, the following step, building value on top of the already-establish hydrodynamic formulation, is to pursue a possible thermodynamic formulation. This was attempted in \cite{Fabbri:2024lhk}, but with weak results. The main reason is that a thermodynamic formulation needs a thorough analysis of the energy-momentum tensor, not done in \cite{Fabbri:2024lhk}.

In the present paper, we will perform such an analysis of the energy-momentum tensor in a full way by employing the so-called $1\!+\!1\!+\!2$ covariant splitting. Inspired by classical fluid mechanics \cite{Ehlers:1961xww}, the covariant splitting allows to investigate the properties of a given geometry in a coordinate-independent and gauge-invariant way. The original development of the covariant splitting is due to Ellis and co-workers \cite{Ellis:1998ct, Clarkson:2002jz}. The basic idea of this method is to define one or two congruences that determine a decomposition of space-time in terms of lower dimensional manifolds called foils.\footnote{Strictly speaking, this is possible only if the congruences are nonvortical. However, covariant approaches extend the notion of foils also to the vortical case by defining some equivalent tensors that allow to treat in a unified way of all the cases \cite{Ellis:1998ct}.} On these foils, one can define a set of tensors (for transformations that preserve the foliation) that characterize the geometry of the flow as well as the thermodynamics of the source fluid. The Bianchi and Ricci identities determine the evolution equations for these variables, which are a closed set of first-order differential equations called $1\!+\!1\!+\!2$ equations. These $1\!+\!1\!+\!2$ covariant equations can be used to formulate a Covariant Gauge Invariant (CGI) theory of perturbations for Locally Rotationally Symmetric (LRS) spacetimes \cite{bed, eb, ehb, ebh}.

Albeit mainly used in astrophysics and cosmology, covariant approaches are a powerful tool that can be used in a general context, provided that a time-like and a space-like congruence are definable. And for the Dirac theory written in hydrodynamic form, these two objects are indeed present. So, they can be used for the $1\!+\!1\!+\!2$ covariant splitting of the energy-momentum tensor, needed to analyze thermodynamic properties. This is what we will do here.

Structure of the paper: in sec. \ref{II} we present the geometrical structure of the polar formulation of spinors and its $1\!+\!1\!+\!2$ covariant splitting; in sec. \ref{III} we will see the same construction for the field equations of the theory, and in particular for the energy tensor; in sec. \ref{IV} we will fix the concepts with a specific example, the Hydrogen Atom; in sec. \ref{V} we will see an application to superconductivity; and in sec. \ref{VI} we will conclude by stretching the theory to investigate an intriguing parallel between the thermodynamics of spinors and the that of a van der Waals gas.
%%%%%%%%%%%%%%%%%%%%%%%%%%%%%%%%%%%%%%%%%%%%%%%%%%%%%%%%%%%%%%%%%%%%%%%%%%%%%%%%%%%%%%%%%%%%%%%%%%%
%%%%%%%%%%%%%%%%%%%%%%%%%%%%%%%%%%%%%%%%%%%%%%%%%%%%%%%%%%%%%%%%%%%%%%%%%%%%%%%%%%%%%%%%%%%%%%%%%%%
\section{Geometry}\label{II}
As conventions, we employ the signature $(+\, -\, -\, -)$. We use Latin indices to indicate coordinate indices, subject to diffeomorphisms, and Greek indices to indicate world indices, subject to real Lorentz transformations (notice that this notation is reverse compared to previous papers \cite{Fabbri:2023onb, Fabbri:2023yhl, Fabbri:2025ftm, Fabbri:2024avj, Fabbri:2024lhk}). For the spinor indices instead, we use the notation for which spinor fields $\psi$ are vectors in spinor space, their adjoint $\overline{\psi}$ are co-vectors in spinor space, and the Clifford matrices $\boldsymbol{\gamma}$ are matrices in spinor space (we also remark that while this is the standard notation, in some cases spinor indices are explicitly indicated with capitalized Latin indices \cite{BJ, Pen}). In addition, because of the existence of frame $e^{a}_{\mu}$ and co-frame $e^{\mu}_{a}$ (such that $e^{a}_{\mu}e^{\mu}_{b}\!=\!\delta^{a}_{b}$ and $e^{a}_{\mu}e^{\nu}_{a}\!=\!\delta^{\nu}_{\mu}$), the passage between Greek and Latin indices can always be performed, and therefore we will retain the right to use Latin or Greek indices whenever we can. This will be true throughout the paper except in the section about the hydrogen atom, where the necessity to make indices explicit will force use to write them in full: in that section, Greek indices are $(t,\ r,\ \theta,\ \varphi)$ for time, radius, elevation angle and azimuthal angle, in spherical coordinates, and Latin indices run over the numerals $(0,\ 1,\ 2,\ 3)$. The background will be taken as that of a general pseudo-Riemannian manifold of metric $g_{ij}$ and we will choose to work with a metric-compatible torsionless Levi-Civita connection only. Finally, the commutation is meant as $[a,b]\!=\!ab\!-\!ba$ whether the objects in brackets are indices of tensors or operators (so we will write $u_{[a}s_{b]}\!=\!u_{a}s_{b}\!-\!u_{b}s_{a}$ as well as $[\nabla_{a},\!\nabla_{b}]T\!=\!\nabla_{a}\nabla_{b}T\!-\!\nabla_{b}\nabla_{a}T$).

\subsection{Algebra}
The Clifford matrices $\boldsymbol{\gamma}^{\mu}$ are such that
\begin{eqnarray}
&\{\boldsymbol{\gamma}^{\mu},\boldsymbol{\gamma}^{\nu}\}\!=\!2\eta^{\mu\nu}\mathbb{I}
\end{eqnarray}
where $\eta^{\mu\nu}$ is the Minkowskian matrix. The Minkowskian matrix and its inverse $\eta_{\mu\nu}$ are used to move indices up and down according to $\boldsymbol{\gamma}^{\mu}\eta_{\mu\nu}\!=\!\boldsymbol{\gamma}_{\nu}$ as usual for tensors. From the Clifford matrices we can define
\begin{eqnarray}
&\boldsymbol{\sigma}_{\mu\nu}\!=\!\frac{1}{4}[\boldsymbol{\gamma}_{\mu},\boldsymbol{\gamma}_{\nu}]
\end{eqnarray}
which are the generators of the Lorentz algebra. With the completely-antisymmetric Levi-Civita pseudo-tensor $\varepsilon_{\mu\nu\rho\sigma}$ it is possible to see that
\begin{eqnarray}
&2i\boldsymbol{\sigma}_{\mu\nu}\!=\!\varepsilon_{\mu\nu\rho\sigma}\boldsymbol{\pi}\boldsymbol{\sigma}^{\rho\sigma}
\end{eqnarray}
implicitly defines the $\boldsymbol{\pi}$ matrix.\footnote{This is usually denoted as $\boldsymbol{\gamma}^{5}$ but as the index $5$ is not a true index, we prefer to use an index-free notation.} The tetrads and co-tetrads are used to exchange Greek to Latin indices as $\boldsymbol{\gamma}_{\mu}e^{\mu}_{i}\!=\!\boldsymbol{\gamma}_{i}$ and back $\boldsymbol{\gamma}^{i}e^{\mu}_{i}\!=\!\boldsymbol{\gamma}^{\mu}$ and of course the metric and its inverse $g^{ij}$ are now used to move indices up and down according to $\boldsymbol{\gamma}_{j}g^{ij}\!=\!\boldsymbol{\gamma}^{i}$ again as can be done for every tensor. Given a spinor $\psi$ and its adjoint\footnote{Notice that because the matrix $\boldsymbol{\gamma}^{0}$ does not transform under spinor transformations, the adjunction is covariantly well-defined.} $\overline{\psi}\!=\!\psi^{\dagger}\boldsymbol{\gamma}^{0}$ we can form spinorial bi-linears according to
\begin{eqnarray}
&K^{ab}\!=\!2\overline{\psi}\boldsymbol{\sigma}^{ab}\boldsymbol{\pi}\psi\ \ \ \
\ \ \ \ \ \ \ \ M^{ab}\!=\!2i\overline{\psi}\boldsymbol{\sigma}^{ab}\psi\\
&S^{a}\!=\!\overline{\psi}\boldsymbol{\gamma}^{a}\boldsymbol{\pi}\psi\ \ \ \
\ \ \ \ \ \ \ \ U^{a}\!=\!\overline{\psi}\boldsymbol{\gamma}^{a}\psi\\
&\Theta\!=\!i\overline{\psi}\boldsymbol{\pi}\psi\ \ \ \
\ \ \ \ \ \ \ \ \Phi\!=\!\overline{\psi}\psi
\end{eqnarray}
which are all tensors. We have $K^{ab}\!=\!-\frac{1}{2}\varepsilon^{abij}M_{ij}$ and $M_{ab}(\Theta^{2}\!+\!\Phi^{2})\!=\!\Phi U^{j}S^{k}\varepsilon_{jkab}\!+\!\Theta U_{[a}S_{b]}$ showing that whenever $\Theta^{2}\!+\!\Phi^{2}\!\neq\!0$ then all the bi-linears are writable in terms of the two vector and the two scalar fields. These verify
\begin{eqnarray}
&U_{a}U^{a}\!=\!-S_{a}S^{a}\!=\!\Theta^{2}\!+\!\Phi^{2}\label{NORM}\\
&U_{a}S^{a}\!=\!0\label{ORTHOGONAL}
\end{eqnarray}
for which condition $\Theta^{2}\!+\!\Phi^{2}\!\neq\!0$ translates into $U_{a}U^{a}\!>\!0$ and $S_{a}S^{a}\!<\!0$ telling that $U^{a}$ is time-like while $S_{a}$ is space-like, so that they can be recognized as the velocity density vector and the spin density axial-vector fields. Spinor fields can be written in the so-called \emph{polar form} which, in chiral representation, reads
\begin{eqnarray}
&\psi\!=\!\phi\ e^{-\frac{i}{2}\beta\boldsymbol{\pi}}
\ \boldsymbol{L}^{-1}\left(\begin{tabular}{c}
$1$\\
$0$\\
$1$\\
$0$
\end{tabular}\right)
\label{spinor}
\end{eqnarray}
for a pair of functions $\phi$ and $\beta$ and for some $\boldsymbol{L}$ with the structure of a spinor transformation \cite{jl1}. With it, we get
\begin{eqnarray}
&\Theta\!=\!2\phi^{2}\sin{\beta}\ \ \ \ \ \ \ \ \ \ \ \ \ \ \ \ \Phi\!=\!2\phi^{2}\cos{\beta}\label{scalars}
\end{eqnarray}
showing that $\beta$ and $\phi$ are a pseudo-scalar and a scalar, called chiral angle and density, and we can introduce
\begin{eqnarray}
&S^{a}\!=\!2\phi^{2}s^{a}\ \ \ \ \ \ \ \ \ \ \ \ \ \ \ \ U^{a}\!=\!2\phi^{2}u^{a}\label{vectors}
\end{eqnarray}
as the spin axial-vector field and velocity vector fields. Consequently, (\ref{NORM}-\ref{ORTHOGONAL}) reduce to
\begin{eqnarray}
&u_{a}u^{a}\!=\!-s_{a}s^{a}\!=\!1\\
&u_{a}s^{a}\!=\!0
\end{eqnarray}
showing that the velocity vector field and the spin axial-vector field have indeed all the properties that are necessary for them to be the generators of the two congruences needed for the $1\!+\!1\!+\!2$ covariant splitting.

According to the general presentation of the covariant splitting, we define the projector
\begin{eqnarray}
&N_{ab}\!=\!g_{ab}\!-\!u_{a}u_{b}\!+\!s_{a}s_{b}
\end{eqnarray}
verifying
\begin{eqnarray}
&N_{ab}u^{a}\!=\!N_{ab}s^{a}\!=\!0\ \ \ \ \ \ \ \
N_{ab}N^{ac}\!=\!N_{b}^{c}\ \ \ \ \ \ \ \
\ \ \ \ \ \ \ \ N^{a}_{a}\!=\!2
\end{eqnarray}
as well as
\begin{eqnarray}
\varepsilon_{ab}\!=\!\varepsilon_{abij}u^{i}s^{j}
\end{eqnarray}
verifying
\begin{eqnarray}
&\varepsilon_{ab}u^{a}\!=\!\varepsilon_{ab}s^{a}\!=\!0\ \ \ \ \ \ \ \
\varepsilon_{ab}\varepsilon^{ij}\!=\!N_{a}^{i}N_{b}^{j}
\!-\!N_{b}^{i}N_{a}^{j}\ \ \ \ \ \ \ \
\varepsilon_{ac}\varepsilon^{bc}\!=\!N_{a}^{b}
\ \ \ \ \ \ \ \
\varepsilon_{ab}\varepsilon^{ab}\!=\!2
\end{eqnarray}
as general identities (these definitions are taken from \cite{EE, CLARK}, although specific conventions and notations may vary).

While it is possible from spinors to form bi-linear that are real tensors, the converse is not possible. However, it is always possible from world tensors to move to coordinate tensors and viceversa. In the following we will establish the condition under which such passage can be done also when tensors are covariant derivatives of other tensors.
%%%%%%%%%%%%%%%%%%%%%%%%%%%%%%%%%%%%%%%%%%%%%%%%%%%%%%%%%%%%%%%%%%%%%%%%%%%%%%%%%%%%%%%%%%%%%%%%%%%
\subsection{Differential construction}
The covariant derivative is defined in terms of the (symmetric and metric-compatible) Levi-Civita connection, which in turn is used to define the spin connection $C_{\alpha\nu k}$ so that
\begin{eqnarray}
&\boldsymbol{\nabla}_{k}\psi\!=\partial_{k}\psi
\!+\!\frac{1}{2}C_{\alpha\nu k}\boldsymbol{\sigma}^{\alpha\nu}\psi\!+\!iqA_{k}\psi
\end{eqnarray}
in which the object $qA_{k}$ is a gauge potential of charge $q$ later identifiable with the electrodynamic potential.\footnote{This is in fact the most general structure for the covariant derivative of spinor fields in absence of conformal symmetry, as it was shown for instance in \cite{Fabbri:2017lmf}: it is intriguing that generality arguments allow in the covariant derivative, beside the gravitational effects, only the electrodynamic interaction, which are the only two actions present for the single spinor field.} General properties of the Lie theory ensure that the logarithmic derivative of an element of a Lie group always belongs to its Lie algebra: in particular for the $\boldsymbol{L}$ in (\ref{spinor}), we can write
\begin{eqnarray}
&\boldsymbol{L}^{-1}\partial_{k}\boldsymbol{L}\!=\!iq\partial_{k}\tau\mathbb{I}
\!+\!\frac{1}{2}\partial_{k}\tau_{\alpha\nu}\boldsymbol{\sigma}^{\alpha\nu}\label{spintrans}
\end{eqnarray}
for some $\partial_{k}\tau$ and $\partial_{k}\tau_{\alpha\nu}$ known as the Goldstone fields of the spinor \cite{Fabbri:2021mfc} (and where $q$ was introduced, without loss of generality, for later convenience). We can now define the two objects
\begin{eqnarray}
&\partial_{k}\tau_{\alpha\nu}\!-\!C_{\alpha\nu k}\!\equiv\!R_{\alpha\nu k}\label{R}\\
&q(\partial_{k}\tau\!-\!A_{k})\!\equiv\!P_{k}\label{P}
\end{eqnarray}
which are proven to be a real tensor and a gauge-covariant vector called \emph{tensorial connection and momentum} \cite{Fabbri:2024lhk}. For completeness, it is important to notice that while the spin connection $C_{\alpha\nu k}$ is defined with both Greek and Latin indices, the tensorial connection $R_{\alpha\nu k}$ is a tensor and thus, even if it is defined with both types of indices, one could convert all of its indices into one type only. In particular, in the form $R_{abk}$ with all Latin indices, we do not need any explicit assignment of tetrads to write it down. We will go back to this point when giving the field equations. As we are now equipped with these two objects, we can write
\begin{eqnarray}
&\boldsymbol{\nabla}_{k}\psi\!=\!(\nabla_{k}\ln{\phi}\mathbb{I}
\!-\!\frac{i}{2}\nabla_{k}\beta\boldsymbol{\pi}
\!-\!\frac{1}{2}R_{abk}\boldsymbol{\sigma}^{ab}
\!-\!iP_{k}\mathbb{I})\psi
\label{decspinder}
\end{eqnarray}
for the covariant derivative of spinor fields in polar form. Moreover, we have
\begin{eqnarray}
&\nabla_{k}s_{j}\!=\!s^{i}R_{ijk}\ \ \ \ \ \ \ \
\ \ \ \ \ \ \ \ \nabla_{k}u_{j}\!=\!u^{i}R_{ijk}\label{ds-du}
\end{eqnarray}
as general identities tying the covariant derivatives of the spin and the velocity to the tensorial connection and which can be inverted. In fact, with the help of the $N_{ij}$ and $\varepsilon_{ab}$ tensors, we have
\begin{eqnarray}
&R_{abk}\!=\!u_{a}\nabla_{k}u_{b}\!-\!u_{b}\nabla_{k}u_{a}
\!+\!s_{b}\nabla_{k}s_{a}\!-\!s_{a}\nabla_{k}s_{b}
\!+\!(u_{a}s_{b}-u_{b}s_{a})\nabla_{k}u_{c}s^{c}
\!+\!2\varepsilon_{ab}V_{k}\label{Rfull}
\end{eqnarray}
making the tensorial connection explicitly written in terms of the covariant derivatives of spin and velocity and in terms of a vector $V_{k}$. It is important to remark that this vector $V_{k}$ must be present as the covariant derivatives of spin and velocity cannot encode all information about the spinor field. In fact, in (\ref{spinor}), take $\boldsymbol{L}\!=\!\mathbb{I}$, corresponding to the fact that the spinor field is in its rest-frame with spin aligned along the third axis: here, rotations around the third axis can have no effect on the velocity (whose spatial part is zero) and no effect on the spin (by construction), and so they can have no impact on their covariant derivatives. Yet, they do have an impact on the spinor field itself, and as a consequence they must be encoded within the covariant derivative of the spinor field. This means that rotations around the spin axis must be encoded either in $P_{k}$ or in the part of $R_{abk}$ that is not given by the covariant derivatives of velocity and spin, which is $V_{k}$. Indeed, we will see that these rotations are encoded in both, and that it is only the difference $P_{k}\!-\!V_{k}$ that has physical significance. Because $P_{k}$ is the momentum of the matter distribution, $P_{k}\!-\!V_{k}$ has to be recognized as what we can call the effective momentum of the material distribution.

The directional derivatives will be denoted as
\begin{eqnarray}
&u^{i}\nabla_{i}\ln{\phi^{2}}\!=\!(\ln{\phi^{2}})\dot \ \ \ \ \ \ \ \
\ \ \ \ \ \ \ \ s^{i}\nabla_{i}\ln{\phi^{2}}\!=\!(\ln{\phi^{2}})\hat \ \ \ \ \ \ \ \
\ \ \ \ \ \ \ \ N^{i}_{a}\nabla_{i}\ln{\phi^{2}}\!=\!\delta_{a}\ln{\phi^{2}}\\
&\ \ \ \ \ \ \ \ u^{i}\nabla_{i}\beta\!=\!\dot{\beta}\ \ \ \ \ \ \ \
\ \ \ \ \ \ \ \ s^{i}\nabla_{i}\beta\!=\!\hat{\beta}\ \ \ \ \ \ \ \
\ \ \ \ \ \ \ \ N^{i}_{a}\nabla_{i}\beta\!=\!\delta_{a}\beta;
\end{eqnarray}
as for the other quantities, we have the scalars
\begin{eqnarray}
&\!\!\!\!\!\!\nabla_{i}u^{i}\!=\!\theta\ \ \ \ \ \
\frac{1}{3}(N^{ij}+2s^{i}s^{j})\nabla_{i}u_{j}\!=\!\Sigma\ \ \ \ \ \
\frac{1}{2}\nabla_{a}u_{b}\varepsilon^{ab}\!=\!\Omega\ \ \ \ \ \
s^{a}u^{b}\nabla_{b}u_{a}\!=\!\mathcal{A}\ \ \ \ \ \ \ \
N^{ab}\nabla_{a}s_{b}\!=\!\varphi\ \ \ \ \ \
\frac{1}{2}\nabla_{a}s_{b}\varepsilon^{ab}\!=\!\xi
\end{eqnarray}
the vectors
\begin{eqnarray}
&\!\!\!\!\!\!\!\!\frac{1}{2}N^{ai}s^{j}(\nabla_{i}u_{j}\!+\!\nabla_{j}u_{i})\!=\!\Sigma^{a}\ \ \ \
\frac{1}{2}N_{ab}\varepsilon^{bijk}u_{i}\nabla_{j}u_{k}\!=\!\Omega_{a}\ \ \ \
N^{ia}u^{b}\nabla_{b}u_{a}\!=\!\mathcal{A}^{i}\ \ \ \ \ \ \ \
N^{ia}s^{b}\nabla_{b}s_{a}\!=\!a^{i}\ \ \ \
N^{ia}u^{b}\nabla_{b}s_{a}\!=\!\alpha^{i}
\end{eqnarray}
and the symmetric irreducible tensors
\begin{eqnarray}
&\frac{1}{2}(N^{j}_{a}N_{b}^{k}\!+\!N^{j}_{b}N_{a}^{k}\!-\!N_{ab}N^{kj})\nabla_{j}u_{k}
\!=\!\Sigma_{ab}\ \ \ \ \ \ \ \ \ \ \ \ \ \ \ \
\frac{1}{2}(N^{j}_{a}N_{b}^{k}\!+\!N^{j}_{b}N_{a}^{k}\!-\!N_{ab}N^{kj})\nabla_{j}s_{k}
\!=\!\zeta_{ab}:
\end{eqnarray}
with all these definitions we can decompose
\begin{eqnarray}
\nonumber
&\nabla_{i}u_{j}\!=\!\Sigma_{ij}
-(\Sigma_{i}s_{j}+\Sigma_{j}s_{i})
+\frac{1}{2}\Sigma(N_{ij}+2s_{i}s_{j})-\\
&-s_{[i}\varepsilon_{j]c}\Omega^{c}
+\varepsilon_{ij}\Omega
+u_{i}\mathcal{A}_{j}
-\mathcal{A}u_{i}s_{j}
+\frac{1}{3}\theta(N_{ij}\!-\!s_{i}s_{j})\\
\nonumber
&\nabla_{i}s_{j}\!=\!\zeta_{ij}
-s_{i}a_{j}
+(\Sigma\!-\!\frac{1}{3}\theta)s_{i}u_{j}
-\Sigma_{i}u_{j}+\\
&+\varepsilon_{ic}\Omega^{c}u_{j}
-\mathcal{A}u_{i}u_{j}
+u_{i}\alpha_{j}
+\varepsilon_{ij}\xi
+\frac{1}{2}N_{ij}\varphi
\end{eqnarray}
in general. Then (\ref{Rfull}) becomes
\begin{eqnarray}
\nonumber
&R_{abk}\!=\!u_{[a}\Sigma_{b]k}
\!+\!u_{[a}\mathcal{A}_{b]}u_{k}
\!-\!u_{[a}\varepsilon_{b]k}\Omega
\!+\!(\frac{1}{3}\theta\!+\!\frac{1}{2}\Sigma)u_{[a}N_{b]k}-\\
\nonumber
&-s_{[a}\zeta_{b]k}
\!+\!s_{[a}a_{b]}s_{k}
\!+\!s_{[a}\varepsilon_{b]k}\xi
\!-\!\frac{1}{2}\varphi s_{[a}N_{b]k}-\\
\nonumber
&-u_{[a}\Sigma_{b]}s_{k}
\!-\!u_{[a}\varepsilon_{b]c}\Omega^{c}s_{k}
\!-\!s_{[a}\alpha_{b]}u_{k}-\\
&-u_{[a}s_{b]}(\mathcal{A}u_{k}\!+\!\frac{1}{3}\theta s_{k}\!-\!\Sigma s_{k}
\!+\!\Sigma_{k}\!-\!\varepsilon_{kc}\Omega^{c})
\!+\!2\varepsilon_{ab}V_{k}\label{Rwhole}
\end{eqnarray}
which will be useful when we will decompose the dynamical equations. Notice that in this form with all world indices and covariantly split, the tensorial connection is in the form that is least dependent on the coordinate system.
%%%%%%%%%%%%%%%%%%%%%%%%%%%%%%%%%%%%%%%%%%%%%%%%%%%%%%%%%%%%%%%%%%%%%%%%%%%%%%%%%%%%%%%%%%%%%%%%%%%
%%%%%%%%%%%%%%%%%%%%%%%%%%%%%%%%%%%%%%%%%%%%%%%%%%%%%%%%%%%%%%%%%%%%%%%%%%%%%%%%%%%%%%%%%%%%%%%%%%%
\section{Dynamical Equations}\label{III}
\subsection{Hydrodynamic Form}
Having collected the definitions of all the relevant geometrical objects, we may next proceed to analyze the dynamics. The dynamical character of the relativistic spinor field theory is assigned by the Dirac equation
\begin{eqnarray}
&i\boldsymbol{\gamma}^{k}\boldsymbol{\nabla}_{k}\psi\!-\!m\psi\!=\!0
\label{D}
\end{eqnarray}
whose polar form can be obtained by first substituting the covariant derivative with (\ref{decspinder}). After this, the result can be multiplied on the left by $\overline{\psi}$, $\overline{\psi}\boldsymbol{\gamma}^{a}$, $\overline{\psi}\boldsymbol{\sigma}^{ab}$, $\overline{\psi}\boldsymbol{\gamma}^{a}\boldsymbol{\pi}$, $\overline{\psi}\boldsymbol{\pi}$, and in each case, split in real and imaginary parts, yielding ten real tensor equations that can be grouped as
\begin{eqnarray}
&\nabla_{a}\Phi\!-\!B_{a}\Theta\!+\!R_{a}\Phi\!+\!2P^{i}M_{ia}\!=\!0 \label{polvi}\\
&\nabla_{a}\Theta\!+\!B_{a}\Phi\!+\!R_{a}\Theta\!-\!2P^{i}K_{ia}\!+\!2mS_{a}\!=0\label{polar}
\end{eqnarray}
\begin{eqnarray}
&\nabla_{i}M^{ia}\!+\!\frac{1}{2}R^{ija}M_{ij}\!-\!2P^{a}\Phi+\!2mU^{a}=\!0\label{polvr}\\
&\nabla^{i}K_{ia}\!+\!\frac{1}{2}R_{ija}K^{ij}\!+\!2P_{a}\Theta\!=\!0\label{polai}
\end{eqnarray}
\begin{eqnarray}
&\nabla_{i}U^{i}\!=\!0\label{poldivU}\\
&(\nabla_{i}\beta\!+\!B_{i})U^{i}\!+\!2P_{i}S^{i}\!=\!0\label{polLodd}\\
&\nabla^{[a}U^{b]}
\!+\!\varepsilon^{abpq}\nabla_{p}\beta U_{q}
\!-\!\frac{1}{2}R^{ij}_{\phantom{ij}p}\varepsilon_{ijqk}
U^{k}\varepsilon^{abpq}\!+\!2\varepsilon^{abpq}P_{p}S_{q}\!-\!2mM^{ab}\!=\!0\label{polcurlU}
\end{eqnarray}
\begin{eqnarray}
&\nabla_{i}S^{i}\!-\!2m\Theta\!=\!0\label{poldivS}\\
&(\nabla_{i}\beta\!+\!B_{i})S^{i}\!+\!2P_{i}U^{i}\!-\!2m\Phi\!=\!0\label{polLeven}\\
&\nabla^{[a}S^{b]}
\!+\!\varepsilon^{abpq}\nabla_{p}\beta S_{q}
\!-\!\frac{1}{2}R^{ij}_{\phantom{ij}p}\varepsilon_{ijqk}
S^{k}\varepsilon^{abpq}\!+\!2\varepsilon^{abpq}P_{p}U_{q}\!=\!0\label{polcurlS}
\end{eqnarray}
in which $R_{ka}^{\phantom{ka}a}\!=\!R_{k}$ and $\varepsilon_{kabc}R^{abc}/2\!=\!B_{k}$ were introduced. Substituting also the bi-linears, and after diagonalization, the above can be translated, respectively, into
\begin{eqnarray}
&F_{i}\!-\!P^{j}\varepsilon_{ij}\!=\!0\label{m}\\
&E_{i}\!-\!P^{j}u_{[j}s_{i]}\!=\!0\label{b}
\end{eqnarray}
\begin{eqnarray}
&F_{i}\varepsilon^{ia}\!+\!E_{i}u^{[i}s^{a]}\!-\!P^{a}\!=\!0
\label{momentum}\\
&F_{i}u^{[i}s^{a]}\!-\!E_{i}\varepsilon^{ia}\!=\!0
\label{complmomentum}
\end{eqnarray}
\begin{eqnarray}
&F_{i}u^{i}\!=\!0\label{A1}\\
&E_{i}u^{i}\!+\!P_{i}s^{i}\!=\!0\label{A2}\\
&\varepsilon^{abij}E_{i}u_{j}\!+\!F^{[a}u^{b]}
\!+\!\varepsilon^{abij}P_{i}s_{j}\!=\!0\label{A3}
\end{eqnarray}
\begin{eqnarray}
&F_{i}s^{i}\!=\!0\label{B1}\\
&E_{i}s^{i}\!+\!P_{i}u^{i}\!=\!0\label{B2}\\
&\varepsilon^{abij}E_{i}s_{j}\!+\!F^{[a}s^{b]}
\!+\!\varepsilon^{abij}P_{i}u_{j}\!=\!0\label{B3}
\end{eqnarray}
in which
\begin{eqnarray}
&E_{i}\!=\!\frac{1}{2}(B_{i}\!+\!\nabla_{i}\beta\!+\!2ms_{i}\cos{\beta})\label{E}\\
&F_{i}\!=\!\frac{1}{2}(R_{i}\!+\!\nabla_{i}\ln{\phi^{2}}\!+\!2ms_{i}\sin{\beta})\label{F}
\end{eqnarray}
were defined for the sake of simplicity: in this form it is a matter of straightforward algebra to prove that each group is equivalent to any other one. It is a little more involved, but still doable, to demonstrate that each group implies the validity of the Dirac equation: to show this, we will have to compute $i\boldsymbol{\gamma}^{k}\boldsymbol{\nabla}_{k}\psi\!-\!m\psi$ and prove that, when equations (\ref{m}-\ref{b}) are taken, the above object is annihilated. In detail, we have that
\begin{eqnarray}
i\boldsymbol{\gamma}^{k}\boldsymbol{\nabla}_{k}\psi\!-\!m\psi
\!=\!(\nabla_{k}\ln{\phi}i\boldsymbol{\gamma}^{k}
\!+\!\frac{1}{2}\nabla_{k}\beta\boldsymbol{\gamma}^{k}\boldsymbol{\pi}
\!-\!\frac{1}{2}R_{abk}i\boldsymbol{\gamma}^{k}\boldsymbol{\sigma}^{ab}
\!+\!P_{k}\boldsymbol{\gamma}^{k}\!-\!m\mathbb{I})\psi
\end{eqnarray}
having used (\ref{decspinder}). With the definition of $\boldsymbol{\sigma}^{ab}$ and the identity $\boldsymbol{\gamma}^{i}\boldsymbol{\gamma}^{j}\boldsymbol{\gamma}^{k}\!=\!\boldsymbol{\gamma}^{i}g^{jk}\!-\!\boldsymbol{\gamma}^{j}g^{ik}\!+\!\boldsymbol{\gamma}^{k}g^{ij}\!+\!i\varepsilon^{ijka}\boldsymbol{\pi}\boldsymbol{\gamma}_{a}$ we get
\begin{eqnarray}
i\boldsymbol{\gamma}^{k}\boldsymbol{\nabla}_{k}\psi\!-\!m\psi
\!=\!\frac{1}{2}(i\nabla_{k}\ln{\phi^{2}}\boldsymbol{\gamma}^{k}
\!+\!\nabla_{k}\beta\boldsymbol{\gamma}^{k}\boldsymbol{\pi}
\!+\!iR_{k}\boldsymbol{\gamma}^{k}
\!+\!B_{k}\boldsymbol{\gamma}^{k}\boldsymbol{\pi}
\!+\!2P_{k}\boldsymbol{\gamma}^{k}
\!-\!2m\mathbb{I})\psi
\end{eqnarray}
and therefore
\begin{eqnarray}
i\boldsymbol{\gamma}^{k}\boldsymbol{\nabla}_{k}\psi\!-\!m\psi
\!=\!(iF_{k}\boldsymbol{\gamma}^{k}
\!+\!E_{k}\boldsymbol{\gamma}^{k}\boldsymbol{\pi}
\!+\!P_{k}\boldsymbol{\gamma}^{k}
\!-\!mis_{k}\sin{\beta}\boldsymbol{\gamma}^{k}
\!-\!ms_{k}\cos{\beta}\boldsymbol{\gamma}^{k}\boldsymbol{\pi}
\!-\!m\mathbb{I})\psi
\end{eqnarray}
after employing (\ref{E}-\ref{F}). Inserting now (\ref{m}-\ref{b}) we obtain
\begin{eqnarray}
i\boldsymbol{\gamma}^{k}\boldsymbol{\nabla}_{k}\psi\!-\!m\psi
\!=\!P^{j}(i\varepsilon_{kj}\boldsymbol{\gamma}^{k}
\!+\!u_{[j}s_{k]}\boldsymbol{\gamma}^{k}\boldsymbol{\pi}
\!+\!\boldsymbol{\gamma}_{j})\psi
\!-\!m(is_{k}\sin{\beta}\boldsymbol{\gamma}^{k}
\!+\!s_{k}\cos{\beta}\boldsymbol{\gamma}^{k}\boldsymbol{\pi}
\!+\!\mathbb{I})\psi\label{Aux}:
\end{eqnarray}
because (see Appendix \ref{app}) from the identities
\begin{eqnarray}
&u_{[a}s_{b]}\boldsymbol{\sigma}^{ab}\boldsymbol{\pi}\psi\!+\!\psi\!=\!0\label{1}\\
&s_{k}\boldsymbol{\gamma}^{k}\boldsymbol{\pi}\psi\!+\!e^{i\beta\boldsymbol{\pi}}\psi\!=\!0\label{2}
\end{eqnarray}
follows the vanishing of the the two parentheses in (\ref{Aux}), we have that $i\boldsymbol{\gamma}^{k}\boldsymbol{\nabla}_{k}\psi\!-\!m\psi\!=\!0$ as wanted. Since (\ref{m}-\ref{b}) are obtained from the Dirac equation and imply the Dirac equation, it follows that the polar formulation of the Dirac theory and the Dirac theory itself are equivalent. More details on this equivalence can be found in \cite{Fabbri:2023yhl}. The equations (\ref{m}-\ref{b}) are in normal form, specifying all derivatives of the two degrees of freedom, and as such the best-suited for a general assessment of the integrability conditions: in fact, by writing them explicitly, they are
\begin{eqnarray}
&\nabla_{a}\beta\!+\!H_{a}\!+\!2ms_{a}\cos{\beta}\!=\!0\\
&\nabla_{a}\ln{\phi^{2}}\!+\!\Xi_{a}\!+\!2ms_{a}\sin{\beta}\!=\!0
\end{eqnarray}
where
\begin{eqnarray}
&B_{a}\!-\!2P^{j}u_{[j}s_{a]}\!=\!H_{a}\label{pseudoext}\\
&R_{a}\!-\!2P^{i}u^{j}s^{k}\varepsilon_{aijk}\!=\!\Xi_{a}\label{ext}
\end{eqnarray}
have been defined. Now, integrability conditions come from the commutativity of the covariant derivatives of the two scalar degrees of freedom, which eventually read
\begin{eqnarray}
&\nabla_{[a}H_{b]}
\!+\!2m\nabla_{[a}s_{b]}\cos{\beta}
\!+\!2mH_{[a}s_{b]}\sin{\beta}=0\\
&\nabla_{[a}\Xi_{b]}
\!+\!2m\nabla_{[a}s_{b]}\sin{\beta}
\!-\!2mH_{[a}s_{b]}\cos{\beta}=0
\end{eqnarray}
and they must be verified, if solutions are to be found. Notice that in particular, they yield
\begin{eqnarray}
&\nabla_{a}H_{b}\varepsilon^{ab}\!+\!4m\xi\cos{\beta}=0\\
&\nabla_{a}\Xi_{b}\varepsilon^{ab}\!+\!4m\xi\sin{\beta}=0
\end{eqnarray}
showing that only if $\xi\!=\!0$ can we have integrability conditions in a form involving only the external potentials (\ref{pseudoext}-\ref{ext}): the condition $\xi\!=\!0$, expressing the vanishing of the twist, is a powerful condition, held to be valid in various scenarios, as we are going to see later on, when presenting the example of the hydrogen atom.

Equations (\ref{momentum}-\ref{complmomentum}) instead are naturally ready to be projected for the $1\!+\!1\!+\!2$ splitting. After using (\ref{Rwhole}), we get
\begin{gather}
\theta\!+\!(\ln{\phi^{2}})\, \dot \!=\!0\ \ \ \ \ \ \ \ \ \ \ \
\varphi\!-\!\mathcal{A}\!+\!(\ln{\phi^{2}})\, \hat \!-\!2m\sin{\beta}\!=\!0\ \ \ \ \ \ \ \ \ \ \ \
\alpha^{k}\varepsilon_{ka}\!-\!2\Omega_{a}\!+\!\delta_{a}\beta\!=\!0\\
2(P\!-\!V)_{i}u^{i}\!=\!2m\cos{\beta}\!-\!2\Omega\!-\!\hat{\beta}\ \ \ \ \ \ \ \ \ \ \ \
2(P\!-\!V)_{i}s^{i}\!=\!-2\xi\!-\!\dot{\beta}\ \ \ \ \ \ \ \ \ \ \ \
2(P\!-\!V)_{i}N^{ik}\!=\!(a_{j}\!-\!\mathcal{A}_{j}\!+\!\delta_{j}\ln{\phi^{2}})\varepsilon^{jk}\label{Pproj}
\end{gather}
in which we see that only the difference $(P\!-\!V)_{i}$ is dynamically relevant. And this is precisely what we meant when in the last section we said that only the effective momentum is physically significant. As we already stated, the tensorial connection with all Latin indices need no explicit basis of tetrads to be written. And the same is true for module and chiral angle since they are both scalars. So, no tetrad is needed to write the field equations in polar form. This means that while in the standard form of the Dirac equation (\ref{D}) one need have spinors and tetrads to write it down, in its polar form one only needs the true degrees of freedom of the spinorial system. We shall see an example of this fact and of the fact that only $(P\!-\!V)_{i}$ is dynamically relevant when we will present the hydrogen atom.

As seen from any of the polar decompositions, all equations are fully real and manifestly covariant, and no gamma matrix remains. Therefore, any specific representation of gamma matrices would lead to the same polar form, which is compatible with the fact that the Dirac theory itself is representation-independent.
%%%%%%%%%%%%%%%%%%%%%%%%%%%%%%%%%%%%%%%%%%%%%%%%%%%%%%%%%%%%%%%%%%%%%%%%%%%%%%%%%%%%%%%%%%%%%%%%%%%
\subsection{Thermodynamic Formulation}
With the dynamical equations written in hydrodynamic form, we are now ready to study the energy density tensor in thermodynamic terms. For this, it is necessary to give the two identities
\begin{eqnarray}
&R^{i}_{\phantom{i}j\alpha\beta}\!=\!-(\nabla_{\alpha}R^{i}_{\phantom{i}j\beta}
\!-\!\nabla_{\beta}R^{i}_{\phantom{i}j\alpha}
\!+\!R^{i}_{\phantom{i}k\alpha}R^{k}_{\phantom{k}j\beta}
\!-\!R^{i}_{\phantom{i}k\beta}R^{k}_{\phantom{k}j\alpha})\label{Riemann}\\
&qF_{\alpha\beta}\!=\!-(\nabla_{\alpha}P_{\beta}\!-\!\nabla_{\beta}P_{\alpha})\label{Maxwell}
\end{eqnarray}
showing that tensorial connection and momentum are respectively the covariant potentials of the Riemann curvature and the Maxwell strength \cite{Fabbri:2024avj}. The spinor field has energy-momentum and spin density tensors given by
\begin{eqnarray}
&T^{ab}\!=\!\frac{i}{2}(\overline{\psi}\boldsymbol{\gamma}^{a}\boldsymbol{\nabla}^{b}\psi
\!-\!\boldsymbol{\nabla}^{b}\overline{\psi}\boldsymbol{\gamma}^{a}\psi)\\
&S^{ijk}\!=\!\frac{i}{4}\overline{\psi}\{\boldsymbol{\gamma}^{i},\boldsymbol{\sigma}^{jk}\}\psi
\end{eqnarray}
verifying the coupled conservation laws
\begin{eqnarray}
&\nabla_{k}T^{ki}\!-\!S_{abk}R^{abki}\!+\!J_{k}F^{ki}\!=\!0\\
&\nabla_{k}S^{kij}\!+\!\frac{1}{2}T^{[ij]}\!=\!0
\end{eqnarray}
which are ensured by the validity of the Dirac equation (notice that $S_{abc}R^{abck}\!=\!0$ for the Dirac case --- however, for the moment, we will leave it, because its presence will suggest us what path to follow when we intend to verify the energy conservation law in polar form). We recall that there is also the conservation of the current density vector $\nabla_{i}J^{i}\!=\!0$ but because $J^{i}\!=\!qU^{i}$ this conservation law is equivalent to $\nabla_{i}U^{i}\!=\!0$ which can be derived from the conservation law of the spin, as was demonstrated in \cite{Fabbri:2025ftm}. In hydrodynamic form the energy and spin are
\begin{eqnarray}
&T^{ab}\!=\!P^{b}U^{a}\!+\!\frac{1}{2}\nabla^{b}\beta S^{a}
\!-\!\frac{1}{4}R_{ij}^{\phantom{ij}b}\varepsilon^{aijk}S_{k}\label{energynonsymm}\\
&S_{abc}\!=\!\frac{1}{4}\varepsilon_{abck}S^{k}
\end{eqnarray}
with conservation laws
\begin{eqnarray}
&U^{i}\nabla_{i}P^{a}\!+\!\frac{1}{2}\nabla_{i}(\nabla^{a}\beta S^{i}
\!-\!\frac{1}{2}R_{jk}^{\phantom{jk}a}\varepsilon^{ijkq}S_{q})
\!-\!\frac{1}{4}\varepsilon_{ijkq}S^{q}R^{ijka}\!+\!J_{i}F^{ia}\!=\!0\label{consT}\\
&\varepsilon^{abij}\nabla_{i}S_{j}\!+\!2P^{[b}U^{a]}\!+\!\nabla^{[b}\beta S^{a]}
\!-\!\frac{1}{2}R_{ij}^{\phantom{ij}[b}\varepsilon^{a]ijk}S_{k}\!=\!0\label{consS}
\end{eqnarray}
which are just the Mathisson-Papapetrou-Dixon equations \cite{Fabbri:2024avj}. To see that they are implied by the Dirac equations in polar form, we begin by considering that (\ref{consS}) is just the Hodge dual of (\ref{polcurlS}). Equation (\ref{consT}) instead is at a higher-order differential and so it requires more work. To start with, we perform the derivatives, so that, after using (\ref{Maxwell}), we get
\begin{eqnarray}
&U_{a}\nabla^{b}P^{a}
\!+\!m\Theta\nabla^{b}\beta
\!+\!\frac{1}{2}S_{a}\nabla^{a}\nabla^{b}\beta
\!-\!\frac{1}{4}R_{ij}^{\phantom{ij}b}\varepsilon^{ijpq}\nabla_{p}S_{q}
\!-\!\frac{1}{4}\nabla_{a}R_{ij}^{\phantom{ij}b}\varepsilon^{aijk}S_{k}
\!-\!\frac{1}{4}\varepsilon_{aijk}S^{k}R^{aijb}\!=\!0
\end{eqnarray}
in which also (\ref{poldivS}) has been used. Replacing the covariant derivative of the spin axial-vector with (\ref{consS}) gives
\begin{eqnarray}
&U_{a}\nabla^{b}P^{a}
\!+\!m\Theta\nabla^{b}\beta
\!+\!\frac{1}{2}S_{a}\nabla^{a}\nabla^{b}\beta
\!+\!P_{j}U_{i}R^{ijb}
\!+\!\frac{1}{2}\nabla_{j}\beta S_{i}R^{ijb}
\!+\!\frac{1}{2}(\nabla^{b}B^{i}\!-\!B_{a}R^{aib})S_{i}\!=\!0
\end{eqnarray}
after having used (\ref{Riemann}) too. The above is equivalent to the simpler
\begin{eqnarray}
&\nabla^{b}(u_{a}P^{a}\!-\!m\cos{\beta})
\!+\!\frac{1}{2}s_{a}\nabla^{a}\nabla^{b}\beta
\!+\!\frac{1}{2}\nabla_{a}\beta\nabla^{b}s^{a}
\!+\!\frac{1}{2}\nabla^{b}B^{a}s_{a}
\!+\!\frac{1}{2}B_{a}\nabla^{b}s^{a}\!=\!0\label{aux1}
\end{eqnarray}
in which also identities (\ref{ds-du}) have been used. Equation (\ref{aux1}) can be written also as
\begin{eqnarray}
&\nabla^{b}(u_{a}P^{a}\!-\!m\cos{\beta}\!+\!\frac{1}{2}s_{a}\nabla^{a}\beta
\!+\!\frac{1}{2}B^{a}s_{a})\!=\!0
\end{eqnarray}
and because of (\ref{polLeven}) we see that it is verified indeed. This proves that the group (\ref{B1}-\ref{B2}-\ref{B3}) implies both conservation laws. As we anticipated, we have $S_{abc}R^{abck}\!=\!0$ for the Dirac case. Notice that because $\nabla_{i}\nabla_{j}S^{ijk}\!=\!0$ then $\nabla_{i}T^{[ij]}\!=\!0$ and therefore we can write
\begin{eqnarray}
&\nabla_{a}[\frac{1}{2}(T^{ab}\!+\!T^{ba})]\!+\!J_{a}F^{ab}\!=\!0\label{aux2}
\end{eqnarray}
showing that the same conservation law holds also for the symmetric part of the energy (this is the so-called Belinfante procedure). We conclude by remarking that the term in the electrodynamic field can be written, by using the Maxwell equations, as the divergence of a symmetric tensor, so that (\ref{aux2}) is equivalent to
\begin{eqnarray}
&\nabla_{a}[\frac{1}{2}(T^{ab}\!+\!T^{ba})
\!+\!\frac{1}{4}F^{2}g^{ab}\!-\!F^{ai}F^{b}_{\phantom{b}i}]\!=\!0
\end{eqnarray}
in general. From now on, we will focus only on the symmetric part and in the case of no electrodynamics.

A symmetric energy density tensor can be decomposed according to
\begin{eqnarray}
\label{T}
&T_{ab}\!=\!\mu u_{a}u_{b}\!-\!p(N_{ab}\!-\!s_{a}s_{b})
\!+\!\frac{1}{2}\Pi(N_{ab}\!+\!2s_{a}s_{b})
\!+\!(\Pi_{a}s_{b}\!+\!\Pi_{b}s_{a})\!+\!\Pi_{ab}
\!+\!Q(s_{a}u_{b}\!+\!s_{b}u_{a})\!+\!(Q_{a}u_{b}\!+\!Q_{b}u_{a})
\end{eqnarray}
in terms of the projected quantities
\begin{eqnarray}
\mu\!=\!T_{ab}u^{a}u^{b}\ \ \ \ \ \ \ \ \ \ \ \
p\!=\!-\frac{1}{3}T_{ab}(N^{ab}\!-\!s^{a}s^{b})\ \ \ \ \ \ \ \ \ \ \ \
Q\!=\!-T_{ab}s^{a}u^{b}\ \ \ \ \ \ \ \ \ \ \ \
&\Pi\!=\!\frac{1}{3}T_{ab}(N^{ab}\!+\!2s^{a}s^{b})\label{Tscal}\\
Q^{a}\!=\!T_{cd}N^{ca}u^{d}\ \ \ \ \ \ \ \ \ \ \ \
&\Pi^{a}\!=\!-T_{cd}N^{ca}s^{d}\label{Tvec}\\
&\Pi^{ab}\!=\!\left(N^{ac}N^{bd}\!-\!\frac{1}{2}N^{ab}N^{cd}\right)T_{cd}\label{Ttens}
\end{eqnarray}
all of which having a thermodynamic interpretation. When in (\ref{Tscal}-\ref{Ttens}) we plug the symmetric part of (\ref{energynonsymm}), after having substituted the tensorial connection with (\ref{Rfull}), as well as the momentum with (\ref{Pproj}), we get the expression of
\begin{align}
\mu\!=\!2\phi^{2}(m\cos{\beta}\!-\!\Omega\!-\!\hat{\beta}/2)\label{MuPsi}\ \ \ \ \ \ \ \ \ \ \ \
p\!=\!-\frac{1}{3}\phi^{2}(2\Omega\!+\!\hat{\beta})\ \ \ \ \ \ \ \ \ \ \ \
Q\!=\!\phi^{2}(\xi\!+\!\dot{\beta})\ \ \ \ \ \ \ \ \ \ \ \ \ \ \ \ \ \ \ \ \ \ \ \
\Pi\!=\!\frac{2}{3}\phi^{2}(\Omega\!-\!\hat{\beta})\\
Q^{a}\!=\!\frac{1}{2}\phi^{2}\varepsilon^{ak}(2\mathcal{A}_{k}\!-\!a_{k}\!-\!\delta_{k}\ln{\phi^{2}})\ \ \ \ \ \ \ \
\Pi^{a}\!=\!\frac{1}{2}\phi^{2}(\Sigma_{k}\varepsilon^{ka}\!+\!\Omega^{a}\!+\!\delta^{a}\beta)\\
\Pi^{ab}\!=\!-\frac{1}{2}\phi^{2}(\Sigma^{a}_{\phantom{a}j}\varepsilon^{jb}
\!+\!\Sigma^{b}_{\phantom{b}j}\varepsilon^{ja})
\end{align}
which are the thermodynamic components of the energy density tensor expressed in terms of the fundamental variables of the covariant formalism. It is worth looking at the details of these quantities:
\begin{enumerate}
\item The quantity $\mu$ describes the \emph{internal energy} of the effective fluid representing the spinor field, and it is associated with its gravitational mass. Such mass, however, can only be calculated easily in the case of asymptotically flat spacetimes. Despite these difficulties, it is striking to notice that the inertial mass $m$ of the spinor field is not necessarily the same as the mass of its effective fluid counterpart and, consequently, its gravitational mass: the structure of this equation reveals that the chiral angle plays an important role in the gravitational action of the spinor field, and such action is corrected by the vorticity of the spacetime.

\item The quantity $p$ represents the isotropic part of the pressure of the effective spinor fluid. Differently from the energy density, the pressure is entirely generated by the chiral angle (and corrected by the vorticity of the spacetime). Notice also that the trace of the stress-energy tensor $T_{\mu\nu}$ reads
\begin{gather}
\label{trace_T_Psi}
\mathfrak{m}\!=\!\mu\!-\!3p\!=\!2\phi^{2}m\cos{\beta}:
\end{gather}
this quantity is zero when treating the null fluid commonly associated with photons modeled as a null fluid. We see that in this picture, $m$ does not always represent even the inertial mass of the effective spinor fluid and that such inertial term is corrected by the presence of the chiral angle. This form of $\mathfrak{m}$ has led some to speculate that the chiral angle is connected to vacuum polarization \cite{H}.

\item The quantity $\Pi$ represents the scalar part of the shearing pressure and is normally associated with the viscosity of a fluid. We do not expect the spinor field to be intrinsically dissipative, and therefore, this component is assumed just to represent shearing forces in the effective spinor fluid. It is worth remarking on the linear combinations given by
\begin{gather}
\label{pressures}
p_{s}=p\!+\!\Pi\!=\!-\phi^{2} \hat{\beta}\\
p_{\perp}\!=\!\frac{1}{2}\Pi\!-\!p\!=\!\phi^{2}\Omega
\end{gather}
which represent, respectively, the pressure along the direction of the vector $s^k$ and the one orthogonal to it. The $p_s$ depends only on the spatial variation of the chiral angle and can be both positive and negative, i.e., a true pressure or a tension. In the limit $\beta\!\rightarrow\!0$ (as we would have in non-relativistic approximations, for instance), pressure and anisotropic pressure must be opposite, and the effective spinor fluid would have only shearing pressure. This form of pressure could be relevant in gravitational systems, which are highly symmetric, as in the case of spherically symmetric collapse. Notice also that in the case of more than one spinor, say two for simplicity, this pressure term would become zero if the spins of these fields are antiparallel. The orthogonal component of the pressure $p_\perp$, instead, is completely determined by the geometry of the spacetime. In the case of no vorticity $\Omega=0$, the effective spinor fluid will have only a pressure along the spin direction and it will be equal to $3p$. Finally, notice that the first equation in \eqref{MuPsi} can be written as
\begin{equation}
\mu\!=\!\mathfrak{m}\!+\!p_{s}\!-\!2p_{\perp}
\end{equation}
which shows that the gravitational mass of the effective spinor fluid is composed of its inertial mass plus some pressure terms, in line with the well-known fact that, in Einstein gravity, pressure exerts gravitational pull.

\item The quantity $Q$ represents the part of the matter-energy flux that is parallel to the spin vector. It is proportional to the time derivative of the chiral angle, corrected by the twist, and therefore, it is present only in cases in which the underlying spacetime is dynamic. As already said, since we have no reason to think that the effective spinor fluid is intrinsically dissipative, $Q$ cannot be ascribed to real heat exchange, but its presence rather indicates that the frame $u^k$ we have chosen is not a true rest frame for the spinor field. This is an interesting result as in choosing the vector $u^k$, we have aligned this frame with the velocity density vector for the spinor field, and therefore, there should be no fluxes. One way to interpret the presence of this term is that the vector $u^k$ does not define the ``true'' rest frame of the spinor field, but rather that such frame does not take into account the internal degrees of freedom of the field. As in the hydrodynamic representation, there is no intrinsic difference between internal and external degrees of freedom; the latter are viewed as ``motions'' of the field.

\item The quantity $Q^{a}$ represents the component of the matter-energy flux orthogonal to $u^k$ and $s^k$. It is primarily generated by the variation of the density $\phi$ in the directions orthogonal to $u^k$ and $s^k$, and it is corrected by the geometry of the spacetime via the acceleration vector for the timelike and spacelike congruences as well as the vectorial part of the shear and the vorticity. This quantity is generally important in the context of axisymmetric problems, and it is directly related to the rotation of the field along a given axis.

\item The quantity $\Pi^{a}$ represents the vector component of the shearing pressure. It is orthogonal to $u^k$. It is directly related to the variation of the chiral angle orthogonal to $u^k$ and $s^k$, and it is corrected by the shear and vorticity vector. It also plays a role in axisymmetric problems, but unlike $Q_a$, it can also appear in stationary spacetimes.

\item The quantity $\Pi^{ab}$ represents the components of the shearing pressure orthogonal to $u^k$ and $s^k$. As for the vector $\Pi^{a}$, this quantity is present when spherical symmetry is violated.
\end{enumerate}
%%%%%%%%%%%%%%%%%%%%%%%%%%%%%%%%%%%%%%%%%%%%%%%%%%%%%%%%%%%%%%%%%%%%%%%%%%%%%%%%%%%%%%%%%%%%%%%%%%%
%%%%%%%%%%%%%%%%%%%%%%%%%%%%%%%%%%%%%%%%%%%%%%%%%%%%%%%%%%%%%%%%%%%%%%%%%%%%%%%%%%%%%%%%%%%%%%%%%%%
\section{Hydrogen Ground State}\label{IV}
In this section, since we will be looking for explicit solutions to the Dirac equation, the coordinate indices will no longer be labelled by Latin indices but with $(t,\ r,\ \theta,\ \varphi)$ for the temporal coordinate, the radial coordinate, the elevation angle and the azimuthal angle, respectively. Toward the end, we will need to give the tetrad fields, whose indices will be both coordinate and Lorentz indices: these last indices will be labelled with the numerals $(0,\ 1,\ 2,\ 3)$.

For the hydrogen atom, the $1S$ orbital is the least-energy solution of the Dirac equation.

We will work in a flat space-time, for which the metric is
\begin{eqnarray}
&g_{tt}\!=\!1\ \ \ \ \ \ \ \ g_{rr}\!=\!-1\ \ \ \ \ \ \ \ g_{\theta\theta}\!=\!-r^{2}\ \ \ \ \ \ \ \ g_{\varphi\varphi}\!=\!-r^{2}|\!\sin{\theta}|^{2}:
\end{eqnarray}
this generates the Levi-Civita symmetric connection
\begin{eqnarray}
&\Lambda^{r}_{\theta\theta}\!=\!-r
\ \ \ \ \ \ \ \ \Lambda^{r}_{\varphi\varphi}\!=\!-r|\!\sin{\theta}|^{2}\ \ \ \ \ \ \ \ \Lambda^{\theta}_{\theta r}\!=\!\Lambda^{\varphi}_{\varphi r}\!=\!\frac{1}{r}\ \ \ \ \ \ \ \
\Lambda^{\varphi}_{\varphi\theta}\!=\!\cot{\theta}\ \ \ \ \ \ \ \
\Lambda^{\theta}_{\varphi\varphi}\!=\!-\cos{\theta}\sin{\theta}
\end{eqnarray}
as known. Setting $\Gamma^{2}\!=\!1\!-\!\alpha^{2}$ where $\alpha$ is the fine-structure constant, we can introduce
\begin{eqnarray}
\Delta\!=\!\frac{1}{\sqrt{1-\alpha^{2}|\!\sin{\theta}|^{2}}}
\end{eqnarray}
so that we can write
\begin{eqnarray}
&s_{r}=\!-\Delta\cos{\theta}\ \ \ \ \ \ \ \ s_{\theta}\!=\!\Gamma\Delta r\sin{\theta}\label{sols}\\
&u_{t}=\!\Delta\ \ \ \ \ \ \ \ u_{\varphi}\!=\!-\alpha\Delta r(\sin{\theta})^{2}\label{solu}
\end{eqnarray}
as spin and velocity, and thus we have
\begin{align}
&N^{tt}=-\alpha^{2}\Delta^{2}(\sin{\theta})^{2}\ \ \ \ \ \ \ \
N^{rr}=-\Gamma^{2}\Delta^{2}(\sin{\theta})^{2}\ \ \ \ \ \ \ \
N^{\theta\theta}=-\Delta^{2}(\cos{\theta})^{2}/r^{2}\ \ \ \ \ \ \ \
N^{\varphi\varphi}=-\Delta^{2}/r^{2}/(\sin{\theta})^{2}\\
&\ \ \ \ \ \ \ \ \ \ \ \ \ \ \ \ \ \ \ \ \ \ \ \ \ \ \ \ \ \ \ \ \ \ \ \ \ \ \ \
N^{t\varphi}=-\alpha\Delta^{2}/r\ \ \ \ \ \ \ \
N^{r\theta}=-\Gamma\Delta^{2}\cos{\theta}\sin{\theta}/r
\end{align}
together with
\begin{align}
&\varepsilon^{tr}\!=\!-\alpha\Gamma\Delta^{2}(\sin{\theta})^{2}\ \ \ \ \ \ \ \
\varepsilon^{t\theta}\!=\!-\alpha\Delta^{2}\sin{\theta}\cos{\theta}/r\ \ \ \ \ \ \ \
\varepsilon^{r\varphi}\!=\!\Gamma\Delta^{2}/r\ \ \ \ \ \ \ \
\varepsilon^{\theta\varphi}\!=\!\Delta^{2}\cot{\theta}/r^{2}
\end{align}
as the covariant objects built from the metric: with the connection we can compute
\begin{eqnarray}
&\nabla_{\theta}s_{r}\!=\!\Gamma\Delta\sin{\theta}(\Gamma\Delta^{2}\!-\!1)\ \ \ \ \ \ \ \
\nabla_{\theta}s_{\theta}\!=\!r\Delta\cos{\theta}(\Gamma\Delta^{2}\!-\!1)\ \ \ \ \ \ \ \
\nabla_{\varphi}s_{\varphi}\!=\!(\Gamma\!-\!1)r\Delta\cos{\theta}(\sin{\theta})^{2}\\
&\!\!\!\!\!\!\!\nabla_{\theta}u_{t}\!=\!\alpha^{2}\Delta^{3}\sin{\theta}\cos{\theta}\ \ \ \ \ \ \ \
\nabla_{\theta}u_{\varphi}\!=\!-\alpha r\Delta^{3}\cos{\theta}\sin{\theta}\ \ \ \ \ \ \ \
\nabla_{\varphi}u_{r}\!=\!\alpha\Delta(\sin{\theta})^{2}\ \ \ \ \ \ \ \
\nabla_{\varphi}u_{\theta}\!=\!\alpha r\Delta\sin{\theta}\cos{\theta}
\end{eqnarray}
from which $\xi\!=\!0$ and
\begin{align}
&\Omega\!=\!-\frac{1}{2}\alpha\Delta^{3}[\Gamma(\sin{\theta})^{2}\!+\!2(\cos{\theta})^{2}]/r
\end{align}
as well as
\begin{align}
&s^{i}\nabla_{i}s_{r}\!=\!
-\Gamma^{2}\Delta^{2}(\sin{\theta})^{2}(\Gamma\Delta^{2}\!-\!1)/r\ \ \ \
\ \ \ \ \ \ \ \ \ \ \ \ s^{i}\nabla_{i}s_{\theta}\!=\!
-\Gamma\Delta^{2}\sin{\theta}\cos{\theta}(\Gamma\Delta^{2}\!-\!1)\\
&s^{i}\nabla_{i}u_{t}\!=\!
-\Gamma\Delta\sin{\theta}\alpha^{2}\Delta^{3}\sin{\theta}\cos{\theta}/r\ \ \ \
\ \ \ \ \ \ \ \ \ \ \ \ s^{i}\nabla_{i}u_{\varphi}\!=\!
\alpha\Gamma\Delta^{4}\cos{\theta}(\sin{\theta})^{2}\\
&\ \ \ \ \ \ \ \ \ \ \ \ \ \ \ \ \ \ \ \ \ \ \ \ \ \ \ \ \ \ \ \ u^{i}\nabla_{i}s_{\varphi}
\!=\!\alpha\Delta^{2}(\Gamma\!-\!1)\cos{\theta}(\sin{\theta})^{2}\\
&\ \ \ \ \ \ \ \ \ \ \ \ u^{i}\nabla_{i}u_{r}\!=\!\alpha^{2}\Delta^{2}(\sin{\theta})^{2}/r\ \ \ \
\ \ \ \ \ \ \ \ \ \ \ \ u^{i}\nabla_{i}u_{\theta}\!=\!\alpha^{2}\Delta^{2}\sin{\theta}\cos{\theta}
\end{align}
and with $\nabla_{\varphi}u_{i}s^{i}\!=\!-\alpha\Delta^{2}(\Gamma\!-\!1)\cos{\theta}(\sin{\theta})^{2}$ as the covariant objects built from metric and connection. Then
\begin{eqnarray}
&R_{t\varphi\theta}\!=\!-\alpha r \sin{\theta}\cos{\theta}\Delta^{2}\ \ \ \ \ \ \ \
R_{r\theta\theta}\!=\!-r(1\!-\!\Gamma\Delta^{2})\ \ \ \ \ \ \ \
R_{r\varphi\varphi}\!=\!-r|\!\sin{\theta}|^{2}\ \ \ \ \ \ \ \
R_{\theta\varphi\varphi}\!=\!-r^{2}\sin{\theta}\cos{\theta}\label{solR}
\end{eqnarray}
is the tensorial connection. And
\begin{eqnarray}
&P_{t}\!=\!m\Gamma\ \ \ \ \ \ \ \ \ \ \ \ \ \ \ \
P_{\varphi}\!=\!-\frac{1}{2}\label{solP}
\end{eqnarray}
is the momentum. With all these elements one can verify that the formula (\ref{Rfull}) is valid for
\begin{eqnarray}
&V_{\varphi}\!=\!-\frac{1}{2}\Delta^{2}[\Gamma(\sin{\theta})^{2}\!+\!(\cos{\theta})^{2}]
\end{eqnarray}
and all other components zero. Consequently
\begin{eqnarray}
&P_{t}\!-\!V_{t}\!=\!m\Gamma\ \ \ \ \ \ \ \ \ \ \ \ \ \ \ \
P_{\varphi}\!-\!V_{\varphi}\!=\!-\frac{1}{2}(\sin{\theta}\Delta)^{2}\Gamma(\Gamma\!-\!1):
\end{eqnarray}
it is important to remark that this is the object that in the free limit $\alpha\!\rightarrow\!0$ would give $P_{\mu}\!-\!V_{\mu}\!\rightarrow\!(m,0)$ as is supposed to be. So it is this difference that corresponds to the actual momentum. Finally, we have that the chiral angle
\begin{eqnarray}
&\beta\!=\!-\arctan{(\frac{\alpha}{\Gamma}\cos{\theta})}\label{solbeta}
\end{eqnarray}
and the module
\begin{eqnarray}
&\phi^{2}\!=\!K^{2}r^{-2(1-\Gamma)}e^{-2\alpha mr}/\Delta\label{solphi}
\end{eqnarray}
are demonstrated to be the solutions to the Dirac equations in presence of Coulomb potential.

We can now compute the stress-energy tensor, starting from the four scalar components given by internal energy, pressure, anisotropic pressure, and flux, as
\begin{gather}
\mu\!=\!2\phi^{2}\left[m\cos{\beta}\!+\!\frac{1}{2}\alpha\Delta^{3}\left[\Gamma(\sin{\theta})^{2}
\!+\!2(\cos{\theta})^{2}\!+\!\Gamma^{2}(\sin{\theta})^{2}\right]/r\right]\\
p\!=\!\frac{1}{3}\alpha\phi^{2}\Delta^{3}\left[\Gamma(\sin{\theta})^{2}\!+\!2(\cos{\theta})^{2}
\!+\!\Gamma^{2}(\sin{\theta})^{2}\right]/r\\
\Pi\!=\!-\frac{1}{3}\alpha\phi^{2}\Delta^{3}\left[\Gamma(\sin{\theta})^{2}\!+\!2(\cos{\theta})^{2}
\!-\!2\Gamma^{2}(\sin{\theta})^{2}\right]/r\\
Q\!=\!0
\end{gather}
from which we can see that there exists a $\Pi$ non-zero while $Q$ vanishes identically. The vector components are
\begin{gather}
\Pi^{r}\!=\!-\phi^{2}\alpha\Gamma^{2}\Delta^{4}\cos{\theta}(\sin{\theta})^{2}/r\\
\Pi^{\theta}\!=\!-\phi^{2}\alpha\Gamma\Delta^{4}(\cos{\theta})^{2}\sin{\theta}/r^{2}
\end{gather}
and
\begin{gather}
Q^{t}\!=\!-\frac{1}{2}\phi^{2}\alpha\Delta^{2}(\sin{\theta})^{2}
[\Gamma(1\!-\!\Gamma)(2\!-\!\Delta^{2})
\!+\!3\alpha^{2}\Delta^{2}(\Gamma|\!\sin{\theta}|^{2}\!+\!|\!\cos{\theta}|^{2})
\!+\!2mr\alpha\Gamma]/r\\
Q^{\varphi}\!=\!-\frac{1}{2}\phi^{2}\Delta^{2}
[\Gamma(1\!-\!\Gamma)(2\!-\!\Delta^{2})
\!+\!3\alpha^{2}\Delta^{2}(\Gamma|\!\sin{\theta}|^{2}\!+\!|\!\cos{\theta}|^{2})
\!+\!2mr\alpha\Gamma]/r^{2}
\end{gather}
none of which generally zero (although both space-like). The tensor components are
\begin{gather}
\!\!\!\!\!\!\!\!\Pi^{rr}\!=\!-\frac{1}{2}\alpha\Gamma^{3}\phi^{2}\Delta^{5}(\sin{\theta})^{4}/r\ \ \ \ \ \ \ \
\Pi^{r\theta}\!=\!-\frac{1}{2}\alpha\Gamma^{2}\Delta^{5}\phi^{2}(\sin{\theta})^{3}
\cos{\theta}/r^{2}\ \ \ \ \ \ \ \
\Pi^{\theta\theta}\!=\!-\frac{1}{2}\alpha\Gamma\Delta^{5}\phi^{2}(\sin{\theta})^{2} (\cos{\theta})^{2}/r^{3}\\
\Pi^{tt}\!=\!\frac{1}{2}\alpha^{3}\Gamma\phi^{2}\Delta^{5}(\sin{\theta})^{4}/r\ \ \ \ \ \ \ \
\Pi^{t\varphi}\!=\!\frac{1}{2}\alpha^{2}\Gamma\Delta^{5}\phi^{2}(\sin{\theta})^{2}/r^{2}\ \ \ \ \ \ \ \ \Pi^{\varphi\varphi}\!=\!\frac{1}{2}\alpha\Gamma\phi^{2}\Delta^{5}/r^{3}
\end{gather}
also not zero.

As already said, (\ref{sols}-\ref{solu}), (\ref{solR}-\ref{solP}) and (\ref{solbeta}-\ref{solphi}) solve the Dirac equations in polar form. The information contained in (\ref{sols}-\ref{solu}), (\ref{solR}-\ref{solP}) and (\ref{solbeta}-\ref{solphi}) can be re-converted into the usual variables given by the tetrads
\begin{eqnarray}
&\!\!\!\!e_{0}^{t}\!=\!1\\
&\!\!\!\!e_{1}^{r}\!=\!\sin{\theta}\cos{\varphi}\ \ \ \
e_{2}^{r}\!=\!\sin{\theta}\sin{\varphi}\ \ \ \
e_{3}^{r}\!=\!\cos{\theta}\\
&\!\!\!\!e_{1}^{\theta}\!=\!\frac{1}{r}\cos{\theta}\cos{\varphi}\ \ \
e_{2}^{\theta}\!=\!\frac{1}{r}\cos{\theta}\sin{\varphi}\ \ \
e_{3}^{\theta}\!=\!-\frac{1}{r}\sin{\theta}\\
&\!\!\!\!e_{1}^{\varphi}\!=\!-\frac{1}{r\sin{\theta}}\sin{\varphi}\ \ \ \
e_{2}^{\varphi}\!=\!\frac{1}{r\sin{\theta}}\cos{\varphi}
\end{eqnarray}
and the spinor field
\begin{eqnarray}
\psi\!=\!\frac{1}{\sqrt{1+\Gamma}}e^{-iEt}r^{\Gamma-1}e^{-\alpha mr}\left(\begin{array}{c}
\!1\!+\!\Gamma\!\\
\!0\!\\
\!i\alpha\cos{\theta}\!\\
\!i\alpha\sin{\theta}e^{i\varphi}\!
\end{array}\right)
\end{eqnarray}
where spinor and gamma matrices are taken now in standard representation. It is straightforward to prove that these tetrads and spinor field verify the Dirac equation with Coulomb potential. This is the form given in textbooks. After a suitable boost along the second axis and rotation around the same axis, the above tetrads can be transformed into
\begin{eqnarray}
&e_{0}^{t}\!=\!\Delta\ \ \ \ \ \ \ \ \ \ \ \ \ \ \ \
e_{0}^{\varphi}\!=\!\frac{1}{r}\alpha\Delta\\
&e_{2}^{t}\!=\!\alpha\sin{\theta}\Delta\ \ \ \ \ \ \ \ \ \ \ \ \ \ \ \
e_{2}^{\varphi}\!=\!\frac{1}{r\sin{\theta}}\Delta\\
&e_{1}^{r}\!=\!\Gamma\sin{\theta}\Delta\ \ \ \ \ \ \ \ \ \ \ \ \ \ \ \
e_{1}^{\theta}\!=\!\frac{1}{r}\cos{\theta}\Delta\\
&e_{3}^{r}\!=\!\cos{\theta}\Delta\ \ \ \ \ \ \ \ \ \ \ \ \ \ \ \
e_{3}^{\theta}\!=\!-\frac{1}{r}\Gamma\sin{\theta}\Delta
\end{eqnarray}
in terms of which the components of velocity and spin become $u_{0}\!=\!1$ and $s_{3}\!=\!-1$ identically. In this basis, the scalar projections of the energy-momentum tensor are, of course, the same. The vector projections are
\begin{gather}
\Pi^{1}\!=\!-\phi^{2}\alpha\Gamma\Delta^{3}\sin{\theta}\cos{\theta}/r\\
Q^{2}\!=\!-\frac{1}{2}\phi^{2}\Delta\sin{\theta}[\Gamma(1\!-\!\Gamma)(2\!-\!\Delta^{2})
\!+\!3\alpha^{2}\Delta^{2}(\Gamma|\!\sin{\theta}|^{2}\!+\!|\!\cos{\theta}|^{2})
\!+\!2mr\alpha\Gamma]/r
\end{gather}
for the anisotropic pressure and flux. The tensor projection is only
\begin{gather}
\Pi^{11}\!=\!-\Pi^{22}\!=\!-\frac{1}{2}\alpha\Gamma\phi^{2}\Delta^{3}(\sin{\theta})^{2}/r
\end{gather}
for the anisotropic pressure. In \cite{BEG} it was reported that the stability of the proton may be due to non-trivial pressure distribution over quarks. Speculations about the internal shear forces acting on the quarks were also discussed. Indeed, concepts like pressure, surface tension, shear, radius, are all part of a recent trend of investigations in which nucleons are treated in terms of mechanical elements \cite{Polyakov:2018zvc, Lorce:2025oot, Lorce:2018egm}. However, these mechanical concepts are not necessarily rooted in the non-trivial internal structure of the nucleon. In fact, also fundamental objects like electrons in outer shells of hydrogen atoms display stability and stress \cite{Freese:2024rkr}. Here we have seen that shear, or anisotropic pressure, as well as heat flux, are present even for particles that do not have any internal structure.
%%%%%%%%%%%%%%%%%%%%%%%%%%%%%%%%%%%%%%%%%%%%%%%%%%%%%%%%%%%%%%%%%%%%%%%%%%%%%%%%%%%%%%%%%%%%%%%%%%%
%%%%%%%%%%%%%%%%%%%%%%%%%%%%%%%%%%%%%%%%%%%%%%%%%%%%%%%%%%%%%%%%%%%%%%%%%%%%%%%%%%%%%%%%%%%%%%%%%%%
\section{Superconductivity}\label{V}
We know from the BCS theory that in a superconductor, within the electronic cloud, individual electrons are bound together into Cooper pairs of opposite spin (bosonization), then behaving collectively as a quasiparticle (condensation): in the process of bosonization, the spin axial-vector is effectively summed to zero. When this happens, we have
\begin{eqnarray}
\mu\!=\!4\phi^{2}P_{b}u^{b}\ \ \ \ \ \ \ \ \ \ \ \
p\!=\!0\ \ \ \ \ \ \ \ \ \ \ \
Q\!=\!-2\phi^{2}P_{a}s^{a}\ \ \ \ \ \ \ \ \ \ \ \
&\Pi\!=\!0\\
Q^{a}\!=\!2\phi^{2}P_{c}N^{ca}\ \ \ \ \ \ \ \ \ \ \ \
&\Pi^{a}\!=\!0\\
&\Pi^{ab}\!=\!0
\end{eqnarray}
showing that all projections about pressures are zero. In particular, all anisotropic pressures are zero, and from fluid dynamics we know that this condition encodes the circumstance of no viscosity \cite{LL}. Taking the momentum with no projection orthogonal to the velocity implies that there is also no heat transfer. Thus, we have adiabaticity.

In this case, after condensation the electronic cloud would result into a superfluid \cite{LL}. In this circumstance, from (\ref{poldivS}) we see that $\Theta\!=\!0$ and so also $\beta\!=\!0$ is valid. As a consequence, in the electronic quasiparticle we get $P^{i}\!=\!mu^{i}$ in general. Because the electric current is defined as $J^{i}\!:=\!qU^{i}\!\equiv\!2q\phi^{2}u^{i}$ then
\begin{eqnarray}
&qnP^{i}\!\equiv\!mJ^{i}\label{eq}
\end{eqnarray}
where $2\phi^{2}\!:=\!n$ as per usual definition is superfluidity: the above is known as constitutive relation. Equation (\ref{eq}) can be explicitly written, via (\ref{P}), according to
\begin{eqnarray}
nq^{2}\nabla_{i}\tau\!-\!nq^{2}A_{i}\!=\!mJ_{i}
\end{eqnarray}
where $\tau$ is the phase of the spinorial wave function: in case of constant phase, it reduces to the London equation; in case of no electrodynamic potential and constant density, it tells that
\begin{eqnarray}
\oint J_{i}dl^{i}\!=\!\frac{q^{2}n}{m}\Delta \tau
\end{eqnarray}
and because the phase difference must always be an integer, it amounts to express the fact that the circulation of the electric current is quantized; if the circulation is zero, it becomes
\begin{eqnarray}
\int\!\!\!\!\int\!\!F_{ij}dS^{ij}\!=\!\Delta \tau
\end{eqnarray}
showing that the electromagnetic flux is also quantized.

Because in superconductivity the electronic quasiparticle is assumed to have constant density, (\ref{Maxwell}) results into
\begin{equation}
m\nabla_{[a}J_{b]}\!=\!-nq^{2}F_{ab}
\label{London}
\end{equation}
which is the London equation for the strength.

The maxwell equations can be worked out to give
\begin{eqnarray}
&\nabla^{2}F^{ai}\!-\!C^{aibk}F_{bk}\!+\!\frac{1}{3}RF^{ai}\!=\!\nabla^{[a}J^{i]}
\end{eqnarray}
where $C_{aibk}$ is the conformal curvature: because of (\ref{London}), we get
\begin{equation}
\nabla^{2}F^{ai}\!-\!C^{aibk}F_{bk}\!+\!\left(\frac{R}{3}\!+\!\frac{q^{2}n}{m}\right)F^{ai}\!=\!0
\end{equation}
which is valid in general. In the case (generally verified) in which the superconductor has no curvature
\begin{equation}
\nabla^{2}F^{ai}\!+\!\left(\sqrt{\frac{q^{2}n}{m}}\right)^{2}F^{ai}\!=\!0
\end{equation}
showing that an effective mass is generated. This is just the inverse of the London penetration depth, and consequently the justification of the Meissner effect. As clear, then, the phenomenology of superconductivity is recovered.

For this, the fact that the condition of zero-average spin translate immediately into the condition of no viscosity is the key element. The covariant splitting of the hydrodynamic form of the Dirac spinor does exactly that.
%%%%%%%%%%%%%%%%%%%%%%%%%%%%%%%%%%%%%%%%%%%%%%%%%%%%%%%%%%%%%%%%%%%%%%%%%%%%%%%%%%%%%%%%%%%%%%%%%%%
%%%%%%%%%%%%%%%%%%%%%%%%%%%%%%%%%%%%%%%%%%%%%%%%%%%%%%%%%%%%%%%%%%%%%%%%%%%%%%%%%%%%%%%%%%%%%%%%%%%
\section{Analogy with van der Waals Gases}\label{VI}
In reference \cite{Fabbri:2024lhk} it was discussed that when the spinor field is in interaction with a torsional background in effective approximation, after introducing $2\phi^{2}\!=\!1/V$, and then $U\!=\!\mu V$, one can manipulate the Dirac equation and the spinor energy to obtain the relations
\begin{gather}
U\!=\!m\cos{\beta}\!+\!3RT\!-\!\frac{a}{V}\ \ \ \ \ \ \ \ \ \ \ \ \ \ \ \ \ \ \ \ \ \ \ \ \ \ \ \
\left(p\!+\!\frac{a}{V^{2}}\right)V\!=\!RT\label{vdW}
\end{gather}
where $a$ is a constant related to torsion and always positive, corresponding to the fact that torsion would be always attractive: for small chiral angle, (\ref{vdW}) would be the internal energy and the equation of state of a van der Waals gas.

The possibility of interpreting the spinor field as a type of van der Waals gas is possible because of the validity of (\ref{vdW}), but in order for these to be obtained from the Dirac theory, it is essential to define the temperature
\begin{eqnarray}
&3RT\!=\!-s^{a}\nabla_{a}\beta/2
\!+\!\frac{1}{2}\varepsilon^{kiab}s_{k}u_{i}\nabla_{a}u_{b}
\end{eqnarray}
where $R$ is the ideal gas constant, introduced to make the comparison clearer. Within a $1\!+\!1\!+\!2$ covariant splitting
\begin{eqnarray}
&3RT\!=\!-\frac{1}{2}\hat{\beta}\!-\!\Omega.
\label{Temp}
\end{eqnarray}

Now, the energy decomposition seen in sec. \ref{III} can be used to compute the energy conditions, which are given, in the strong and weak case respectively, according to
\begin{eqnarray}
&(T^{ab}\!-\!\frac{1}{2}Tg^{ab})u_{a}u_{b}\!\geqslant\!0\ \ \ \ \ \ \ \ \mathrm{and}
\ \ \ \ \ \ \ \ T^{ab}u_{a}u_{b}\!\geqslant\!0.
\end{eqnarray}
After the covariant splitting, the strong and weak energy conditions respectively become
\begin{eqnarray}
&\mu\!+\!3p\!\geqslant\!0\ \ \ \ \ \ \ \ \mathrm{and}\ \ \ \ \ \ \ \ \mu\!\geqslant\!0:
\end{eqnarray}
in the case of the spinor field, we have
\begin{eqnarray}
&m\cos{\beta}\!-\!2\Omega\!-\!\hat{\beta}\!\geqslant\!0\ \ \ \ \ \ \ \ \mathrm{and}
\ \ \ \ \ \ \ \ 2m\cos{\beta}\!-\!2\Omega\!-\!\hat{\beta}\!\geqslant\!0\label{condition}.
\end{eqnarray}

These energy conditions, in the analogy with the van der Waals gas, are just
\begin{eqnarray}
&m\cos{\beta}\!+\!6RT\!\geqslant\!0\ \ \ \ \ \ \ \ \mathrm{and}
\ \ \ \ \ \ \ \ m\cos{\beta}\!+\!3RT\!\geqslant\!0:
\end{eqnarray}
in particular, this shows that if the weak energy condition holds then the strong energy condition holds as well.

For the hydrogen atom, we have $m\cos{\beta}\!=\!m\Gamma\Delta$ as well as
\begin{eqnarray}
&6RT\!=\!\alpha\Delta^{3}[\Gamma(\sin{\theta})^{2}\!+\!\Gamma^{2}(\sin{\theta})^{2}
\!+\!2(\cos{\theta})^{2}]/r
\end{eqnarray}
and therefore the temperature as we just defined is always positive, for the hydrogen atom. In particular, the energy conditions are always verified, at least for hydrogen atoms in ground state.

We conclude with some words of caution: in the above definition the temperature is given in a formal way, in the sense that in terms of (\ref{Temp}) (as well as $2\phi^{2}\!=\!1/V$, and $U\!=\!\mu V$) one can work out the Dirac equation and the spinor energy into the equation of state and the internal energy of a van der Waals gas. Such temperature is defined for one electron, and thus it does \emph{not} represent a chaotic motion of a gas of particles. Consequently, the fact that for the hydrogen atom it turns out to be positive may be accidental. This property is not the reflection of more fundamental principles, as it would be for the standard definition of absolute temperature in thermodynamics. Still, it is built in terms of the chiral angle (which is the phase difference between left and right components of the spinor and, as such, somewhat related to internal degrees of freedom) and the vorticity (a quantity that may be thought as some form of orbital motion): so, as a whole, it is a quantity tied to internal dynamics, as temperature would be. In non-relativistic limit, in fact, it would tend to vanish \cite{Fabbri:2024lhk}. It would therefore be interesting to see if in (\ref{Temp}) the quantity $T$ could be proven to be always positive. In this way, more information about the energy conditions could be inferred.

It is also important to stress that such a definition of temperature, albeit formal, is still given in the classical context of non-relativistic thermodynamics. As such, it may be subject to revision when relativistic effects are accounted for.
%%%%%%%%%%%%%%%%%%%%%%%%%%%%%%%%%%%%%%%%%%%%%%%%%%%%%%%%%%%%%%%%%%%%%%%%%%%%%%%%%%%%%%%%%%%%%%%%%%%
%%%%%%%%%%%%%%%%%%%%%%%%%%%%%%%%%%%%%%%%%%%%%%%%%%%%%%%%%%%%%%%%%%%%%%%%%%%%%%%%%%%%%%%%%%%%%%%%%%%
\section{Conclusion}
In this paper, we have considered the relativistic quantum mechanical theory of spinor fields which, when employing the polar form, can be converted in a type of hydrodynamics. When the $1\!+\!1\!+\!2$ covariant splitting is performed, it is possible to extract from the energy-momentum tensor important thermodynamic properties. When polar decomposition and covariant splitting are taken together, several properties of spinor systems can be described in a cleaner way: in detail, we have computed heat fluxes and pressures within the electronic cloud for the stable orbital in the hydrogen atom, and given general comments on energy conditions; the formal definition of temperature for a single electron has been given in concomitance with the above results, tied to the energy conditions, and computed for the case of the hydrogen atom; some comment about superconductivity was also addressed in a phenomenological context.

With the methods presented here, general treatments of quantum mechanics, which are notoriously difficult, might be taken down to the theory of fluid dynamics, which is somewhat simpler. These results will serve as a basis for the development of covariant approaches to the Dirac field in any of the LRS spacetimes \cite{Vignolo:2025ppo,Vignolo:2026qfn}.

More in general, because spinors naturally couple to the torsion of the background, the most important open issues will be about the spin-torsion mediated interactions: along these lines, avenues of research might span from cold Dark Matter \cite{CapolupoJr} to corrections to the phenomenon of neutrino oscillation and the related $CP$ violation \cite{CapolupoSenior}, to dynamics of inflationary scenarios in early-stage universes \cite{CapolupotheIII}. In these approaches, the treatment of spinors as a condensate would fit within the applicability of the polar formulation and the hydrodynamic methods presented above.
%%%%%%%%%%%%%%%%%%%%%%%%%%%%%%%%%%%%%%%%%%%%%%%%%%%%%%%%%%%%%%%%%%%%%%%%%%%%%%%%%%%%%%%%%%%%%%%%%%%
%%%%%%%%%%%%%%%%%%%%%%%%%%%%%%%%%%%%%%%%%%%%%%%%%%%%%%%%%%%%%%%%%%%%%%%%%%%%%%%%%%%%%%%%%%%%%%%%%%%

\

\textbf{Funding}. This work was funded by the Next Generation EU project ``Geometrical and Topological effects on Quantum Matter (GeTOnQuaM)''.

\

\textbf{Data availability}. There is no data available in a repository.

\

\textbf{Conflict of interest}. The author declares no conflict of interest.
%%%%%%%%%%%%%%%%%%%%%%%%%%%%%%%%%%%%%%%%%%%%%%%%%%%%%%%%%%%%%%%%%%%%%%%%%%%%%%%%%%%%%%%%%%%%%%%%%%%
%%%%%%%%%%%%%%%%%%%%%%%%%%%%%%%%%%%%%%%%%%%%%%%%%%%%%%%%%%%%%%%%%%%%%%%%%%%%%%%%%%%%%%%%%%%%%%%%%%%
\appendix
\section{Some Clifford Identities}\label{app}
The equivalence between the Dirac equation and any of its polar formulations, such as for instance (\ref{m}-\ref{b}), is proven by showing that the Dirac equation implies and is implied by (\ref{m}-\ref{b}): the implication is a direct computation, done now in many reference, such as for example in \cite{Fabbri:2023yhl}; the reverse implication is a more involved check, done in sec. \ref{III} with the help of identities (\ref{1}-\ref{2}). We did not work too much on these since they are just Clifford algebraic identities.

Nevertheless, to fully complete the proof it may be worth elaborating a little on these as well. To do that, we shall demonstrate them. We start from plugging (\ref{spinor}) into (\ref{2}) getting
\begin{eqnarray}
&s_{k}\boldsymbol{\gamma}^{k}\boldsymbol{\pi}\phi\ e^{-\frac{i}{2}\beta\boldsymbol{\pi}}
\ \boldsymbol{L}^{-1}\left(\begin{tabular}{c}
$1$\\
$0$\\
$1$\\
$0$
\end{tabular}\right)\!+\!e^{i\beta\boldsymbol{\pi}}\phi\ e^{-\frac{i}{2}\beta\boldsymbol{\pi}}
\ \boldsymbol{L}^{-1}\left(\begin{tabular}{c}
$1$\\
$0$\\
$1$\\
$0$
\end{tabular}\right)\!=\!0
\end{eqnarray}
and because $[\boldsymbol{\pi},\boldsymbol{L}]\!=\!0$ then checking the validity of the above is equivalent to checking the validity of
\begin{eqnarray}
&s_{k}\boldsymbol{\gamma}^{k}\boldsymbol{\pi}
\ \boldsymbol{L}^{-1}\left(\begin{tabular}{c}
$1$\\
$0$\\
$1$\\
$0$
\end{tabular}\right)\!+\!\ \boldsymbol{L}^{-1}\left(\begin{tabular}{c}
$1$\\
$0$\\
$1$\\
$0$
\end{tabular}\right)\!=\!0:
\end{eqnarray}
because this condition is covariant under spinorial transformations, we can check it by checking its validity in a given frame, for instance the one in which $\boldsymbol{L}$ is the identity. As in this frame $s^{3}\!=\!1$, we get
\begin{eqnarray}
&\boldsymbol{\gamma}^{3}\left(\begin{tabular}{c}
$1$\\
$0$\\
$-1$\\
$0$
\end{tabular}\right)\!+\!\left(\begin{tabular}{c}
$1$\\
$0$\\
$1$\\
$0$
\end{tabular}\right)\!=\!0
\end{eqnarray}
which is easily verified. Thus
\begin{eqnarray}
&s_{k}\boldsymbol{\gamma}^{k}\boldsymbol{\pi}\psi\!+\!e^{i\beta\boldsymbol{\pi}}\psi\!=\!0
\end{eqnarray}
in every frame and for every spinor. With the same reasoning, one could have proved that also
\begin{eqnarray}
&u_{k}\boldsymbol{\gamma}^{k}\psi\!-\!e^{i\beta\boldsymbol{\pi}}\psi\!=\!0
\end{eqnarray}
is a general identity. Repeatedly using these last two expressions allows us to see that
\begin{eqnarray}
&u_{a}\boldsymbol{\gamma}^{a}s_{b}\boldsymbol{\gamma}^{b}\boldsymbol{\pi}\psi
\!-\!s_{a}\boldsymbol{\gamma}^{a}u_{b}\boldsymbol{\gamma}^{b}\boldsymbol{\pi}\psi
\!=\!u_{a}\boldsymbol{\gamma}^{a}(-e^{i\beta\boldsymbol{\pi}}\psi)
\!+\!s_{a}\boldsymbol{\gamma}^{a}\boldsymbol{\pi}(e^{i\beta\boldsymbol{\pi}}\psi)
\!=\!-e^{-i\beta\boldsymbol{\pi}}(e^{i\beta\boldsymbol{\pi}}\psi)
\!+\!e^{-i\beta\boldsymbol{\pi}}(-e^{i\beta\boldsymbol{\pi}}\psi)
\!=\!-2\psi
\end{eqnarray}
which can be written as
\begin{eqnarray}
&u_{[a}s_{b]}\boldsymbol{\gamma}^{a}\boldsymbol{\gamma}^{b}\boldsymbol{\pi}\psi\!+\!2\psi\!=\!0,
\end{eqnarray}
or simply
\begin{eqnarray}
&u_{[a}s_{b]}\boldsymbol{\sigma}^{ab}\boldsymbol{\pi}\psi\!+\!\psi\!=\!0,
\end{eqnarray}
which is just (\ref{1}). Therefore, both (\ref{1}-\ref{2}) are identities, valid for any spinor field.
%%%%%%%%%%%%%%%%%%%%%%%%%%%%%%%%%%%%%%%%%%%%%%%%%%%%%%%%%%%%%%%%%%%%%%%%%%%%%%%%%%%%%%%%%%%%%%%%%%%
%%%%%%%%%%%%%%%%%%%%%%%%%%%%%%%%%%%%%%%%%%%%%%%%%%%%%%%%%%%%%%%%%%%%%%%%%%%%%%%%%%%%%%%%%%%%%%%%%%%

%%%%%%%%%%%%%%%%%%%%%%%%%%%%%%%%%%%%%%%%%%%%%%%%%%%%%%%%%%%%%%%%%%%%%%%%%%%%%%%%%%%%%%%%%%%%%%%%%%%

\begin{thebibliography}{120}
\bibitem{jl1}
G. Jakobi, G. Lochak, ``Introduction des param\`{e}tres relativistes de Cayley-Klein dans la repr\'{e}sentation\\ hydrodynamique de l'\'{e}quation de Dirac'', \textit{Comp. Rend. Acad. Sci.} \textbf{243}, 234 (1956).
\bibitem{t2}
T. Takabayasi, ``Relativistic Hydrodynamics of the Dirac Matter'', \textit{Prog. Theor. Phys. Supplement} \textbf{4}, 1 (1957).
\bibitem{Fabbri:2023onb}
Luca Fabbri, ``Dirac Theory in Hydrodynamic Form'',
\textit{Found. Phys.} \textbf{53}, 54 (2023).
\bibitem{Fabbri:2023yhl}
Luca Fabbri, ``Dirac Hydrodynamics in 19 Forms'',
\textit{Symmetry} \textbf{15}, 1685 (2023).
\bibitem{Fabbri:2025ftm}
Luca Fabbri, ``Madelung structure of the Dirac equation'',
\textit{J. Phys. A"} \textbf{19}, 195301 (2025).
\bibitem{Fabbri:2024avj}
Luca Fabbri, ``Classical characters of spinor fields in torsion gravity'',
\textit{Class. Quant. Grav.} \textbf{24}, 245005 (2024).
\bibitem{Fabbri:2024lhk}
Luca Fabbri, ``Dirac field, van der Waals gas, Weyssenhoff fluid, Newton particle'',
\textit{Foundations} \textbf{4}, 134 (2024).
\bibitem{Ehlers:1961xww}
J. Ehlers, ``Contributions to the relativistic mechanics of continuous media'',\\
\textit{Abh. Akad. Wiss. Lit. Mainz. Nat. Kl.} \textbf{11}, 793 (1961).
\bibitem{Ellis:1998ct}
Ellis, G. F. R., van Elst, H., ``Cosmological models: Cargese lectures 1998'', \textit{NATO Sci. Ser. C} \textbf{541}, 1 (1999).
\bibitem{Clarkson:2002jz}
Clarkson, C. A., Barrett, R. K., ``Covariant perturbations of Schwarzschild\\
black holes'', \textit{Class. Quant. Grav.} \textbf{20}, 3855 (2003).
\bibitem{bed}
M. Bruni, G. F. R. Ellis, P. K. S. Dunsby, ``Gauge-invariant perturbations\\
in a scalar field dominated universe'', \textit{Class. Quant. Grav.} \textbf{9}, 921 (1992).
\bibitem{eb}
Ellis, G. F. R., Bruni, M., ``Covariant and gauge-invariant approach to\\
cosmological density fluctuations'', \textit{Phys. Rev. D} \textbf{40}, 1804 (1989).
\bibitem{ehb}
Ellis, G. F. R., Hwang, J., Bruni, M., ``Covariant and gauge-independent\\
perfect-fluid Robertson-Walker perturbations'', \textit{Phys. Rev. D} \textbf{40}, 1819 (1989).
\bibitem{ebh}
Ellis, G. F. R., Bruni, M., Hwang, J., ``Density-gradient-vorticity relation in\\
perfect-fluid Robertson-Walker perturbations'', \textit{Phys. Rev. D} \textbf{42}, 1035 (1990).
\bibitem{BJ}
W. L. Bade, H. Jehle, ``An introduction to Spinors'',
\textit{Rev. Mod. Phys.} \textbf{25}, 714 (1953).
\bibitem{Pen}
R. Penrose, ``From conformal infinity to equations of motion: conserved quantities\\
in general relativity'', \textit{Phil. Trans. R. Soc. A} \textbf{382}, 20230041 (2024).
\bibitem{EE}
H. van Elst, G. F. R. Ellis, ``The covariant approach to LRS perfect fluid\\
spacetime geometries'', \textit{Class. Quant. Grav.} \textbf{13}, 1099 (1996).
\bibitem{CLARK}
C. Clarkson, ``Covariant approach for perturbations of rotationally symmetric spacetimes'', \textit{Phys Rev. D}, \textbf{76}, 104034 (2007).
\bibitem{Fabbri:2017lmf}
Luca Fabbri, ``Fundamental Theory of Torsion Gravity'', \textit{Universe} \textbf{7}, 305 (2021).
\bibitem{Fabbri:2021mfc}
Luca Fabbri, ``Weyl and Majorana Spinors as Pure Goldstone Bosons'', \textit{Adv. Appl. Clifford Algebras} \textbf{32}, 3 (2022).
\bibitem{H}
D. Hestenes, ``Real Spinor Fields'', \textit{J.Math.Phys.} \textbf{8}, 798 (1967).
\bibitem{BEG}
Burkert, V. D., Elouadrhiri, L., Girod, F. X., ``The pressure distribution inside the proton'', \textit{Nature} \textbf{557}, 396 (2018).
\bibitem{Polyakov:2018zvc}
Polyakov, M. V., Schweitzer, P., ``Forces inside hadrons: pressure, surface tension,\\
mechanical radius, and all that'', \textit{Int. J. Mod. Phys. A} \textbf{33}, 1830025 (2018).
\bibitem{Lorce:2025oot}
Lorc\'{e}, C., Schweitzer, P., ``Pressure inside hadrons: criticism, conjectures,\\
and all that'', \textit{Acta Phys. Polon. B} \textbf{56}, 3--A17 (2025).
\bibitem{Lorce:2018egm}
Lorc\'{e}, C., Moutarde, H., Trawi\'{n}ski, A. P., ``Revisiting the mechanical properties\\
of the nucleon'', \textit{Eur. Phys. J. C} \textbf{79}, 89 (2019).
\bibitem{Freese:2024rkr}
Adam Freese, ``Quantum stresses in the hydrogen atom'',
\textit{Phys. Rev. D} \textbf{111}, 034047 (2025).
\bibitem{LL}
L. D. Landau, E. M. Lifshitz, \textit{Fluid Mechanics} (Butterworth-Heinemann, 1987).
\bibitem{Vignolo:2025ppo}
S. Vignolo, G. De Maria, L. Fabbri, S. Carloni, ``A covariant approach to the Dirac field\\
in LRS space-times'', \textit{Class. Quant. Grav.} \textbf{42}, 215013 (2025).
\bibitem{Vignolo:2026qfn}
S. Vignolo, G. De Maria, S. Carloni, L. Fabbri, ``The Dirac field in LRS space-times:\\
a covariant approach'', arXiv:2605.10620 [gr-qc].
\bibitem{CapolupoJr}
A. Capolupo, S. Capozziello, G. Pisacane, A. Quaranta, ``Missing matter in galaxies\\
as a neutrino mixing effect'', \textit{Phys. Dark Univ.} \textbf{48}, 101894 (2025).
\bibitem{CapolupoSenior}
A. Capolupo, G. De Maria, S. Monda, A. Quaranta, R. Serao, ``Quantum Field Theory\\
of Neutrino Mixing in Spacetimes with Torsion'', \textit{Universe} \textbf{10}, 170 (2024).
\bibitem{CapolupotheIII}
A. Capolupo, S. Carloni, L. Fabbri, S. Monda, A. Quaranta, S. Vignolo, ``Quantized Dirac Fields in torsionful gravity:\\ cosmological implications and links with the dark universe'', \textit{Int. J. Geom. Meth. Mod. Phys.}, to appear.
\end{thebibliography}
\end{document}